\documentclass[11pt]{article}
\usepackage{epsfig} 
\newcommand{\sect}[1]{ \section{#1} \setcounter{equation}{0} }

\newcommand{\pslash}{p \! \! \! /} 
\newcommand{\qslash}{q \! \! \! /}

\newcommand{\third}{\mbox{\small{$\frac{1}{3}$}}} 
 
\newcommand{\pitwo}{\mbox{\small{$\frac{\pi}{2}$}}} 
\newcommand{\pisix}{\mbox{\small{$\frac{\pi}{6}$}}} 
\newcommand{\MSbar}{\overline{\mbox{MS}}} 
\newcommand{\MSbars}{\overline{\mbox{\footnotesize{MS}}}} 
\newcommand{\Nf}{N_{\!f}}

\newcommand{\NA}{N_{\!A}}

\begin{document}
\title{Amplitudes for the $n$~$=$~$3$ moment of the Wilson operator at two 
loops in the RI${}^\prime$/SMOM scheme}
\author{J.A. Gracey, \\ Theoretical Physics Division, \\ 
Department of Mathematical Sciences, \\ University of Liverpool, \\ P.O. Box 
147, \\ Liverpool, \\ L69 3BX, \\ United Kingdom.} 
\date{} 
\maketitle 

\vspace{5cm} 
\noindent 
{\bf Abstract.} The renormalization of the third moment of the twist-$2$
flavour non-singlet Wilson operator is given to two loops in the
RI${}^\prime$/SMOM renormalization scheme. This involves renormalizing the
operator itself and the two total derivative operators into which it mixes by
inserting it into a quark $2$-point function and evaluating the amplitudes at
the symmetric subtraction point. The corresponding two loop conversion 
functions are derived from which the three loop Landau gauge anomalous
dimension is deduced. The full set of amplitudes for the two loop Green's
function for each of the operators are given in both the $\MSbar$ and 
RI${}^\prime$/SMOM schemes.  

\vspace{-18cm}
\hspace{13.5cm}
{\bf LTH 912}

\newpage

\sect{Introduction.}

Recently a renormalization scheme has been introduced which aids the extraction
of accurate measurements of matrix elements or Green's functions from the
lattice for strongly interacting fields, \cite{1,2,3}. The scheme is named
RI${}^\prime$/SMOM which stands for the modified regularization invariant
(RI${}^\prime$) scheme at the symmetric subtraction point where the infinities
are removed in a fashion akin to the momentum subtraction scheme, \cite{4}. The
shorthand for the latter is MOM. Whilst this is a pr\'{e}cis of the syntax it
in effect precisely defines the method for performing the renormalization. The
RI${}^\prime$/SMOM scheme is an extension of the original regularization
invariant (RI) scheme and its modification (RI${}^\prime$), \cite{5,6}. These 
were developed earlier to aid the matching of lattice computations with the 
high energy behaviour determined in conventional perturbation theory. In 
essence those schemes are such that $2$-point functions are rendered finite by 
choosing the renormalization constant so that at the subtraction point there 
are no $O(a)$ corrections where $a$~$=$~$g^2/(16\pi^2)$ and $g$ is the coupling
constant in QCD. By contrast the renormalization of $3$-point functions such as
vertices or operators inserted into $2$-point functions is carried out using
the standard $\MSbar$ scheme, \cite{5,6}. The technical difference between RI 
and RI${}^\prime$ resides in the way the quark $2$-point function is treated 
prior to applying the renormalization condition. More specifically this is how 
to project out the contribution associated with $\pslash$ where $p$ is the 
external momentum of the $2$-point function. The scheme which is more widely
used of the two is RI${}^\prime$ as it minimizes the number of derivatives
required to be taken. This reduces the financial penalty in respect of defining
the scheme when lattice regularization is used. By contrast using dimensional
regularization in the continuum there is no such cost, only the limitation in
the range of validity of the coupling constant, due to its smallness, which is 
not an issue on the lattice. There the nonperturbative structure can be fully 
explored. Further, there have been several continuum computations performed in 
perturbation theory in QCD, \cite{7,8}. For instance, the three and four loop 
renormalization of QCD in RI${}^\prime$ was carried out in the Landau gauge in 
\cite{7} and for an arbitrary linear covariant gauge at three loops in 
\cite{8}. In \cite{8} {\em all} $2$-point functions were made finite in the 
same way as the quark $2$-point function was in the original article, 
\cite{5,6}. Though it should be noted that in the Landau gauge the coupling 
constant and gauge parameter variables are the same in $\MSbar$ and 
RI${}^\prime$. 

We have given a resum\'{e} of the RI${}^\prime$ scheme partly to contrast with
that of RI${}^\prime$/SMOM which is the main focus here, but also since it is
retained as part of the latter scheme. This is because RI${}^\prime$/SMOM 
differs from RI${}^\prime$ in the way $3$-point and higher functions are 
treated. For instance, it was pointed out in \cite{1} that RI${}^\prime$ is
sensitive to infrared effects. This stems from the fact that there is an
exceptional momentum configuration for $3$-point functions and the nullified
momentum flowing through an inserted operator leads to unwanted infrared
singularities. Examples of where one has to be careful in the treatment are the
quark current operators and the Wilson operators used in deep inelastic 
scattering. Therefore, there is a potential problem with making accurate 
measurements of matrix elements involving such operator insertions on
the lattice which is the reason why RI${}^\prime$/SMOM was developed, \cite{1}.
In this scheme there is a momentum flowing through each of the external legs 
and inserted operators of the Green's function in a way which is consistent 
with energy-momentum conservation. Therefore in the absence of an exceptional
momentum configuration there ought to be no infrared problems. Indeed it was
shown in \cite{2,3} that there was an improvement in the convergence of the
conversion function associated with the renormalization at two loops when the
RI${}^\prime$/SMOM scheme was used instead of RI${}^\prime$ for the case of the
scalar and tensor quark currents. More recently the second moment of the 
twist-$2$ flavour non-singlet Wilson operator was studied in \cite{9} at two 
loops. This built on the initial one loop calculation of \cite{10} and the two
loop computations for the quark currents, \cite{2,3,11}. The Wilson operator 
was a more involved computation than the quark currents due to the mixing of 
the operator with a total derivative operator. The latter is completely passive
in the RI${}^\prime$ scheme renormalization as the operator insertion is at 
zero momentum by definition. For RI${}^\prime$/SMOM it cannot be neglected and 
the mixing, which has been studied at three loops in $\MSbar$ in \cite{12}, is 
crucial. Though comparing similar conversion functions for this operator moment
suggested that, if anything, in the RI${}^\prime$ case the perturbative 
convergence of the series was marginally quicker, \cite{9}. Though it was noted
in \cite{9} that as the RI${}^\prime$ scheme does not access the off-diagonal 
part of the mixing matrix, due to the form of the Green's function momentum 
configuration, it is in some sense not a full scheme for operators with mixing,
in contrast to multiplicatively renormalizable operators such as the quark 
currents.

Whilst we have concentrated on the renormalization, one ingredient which is
crucial for lattice computations is the actual structure of the Green's
function with the operator inserted in the quark $2$-point function. This is
used to measure nonperturbative matrix elements. However, in order to assist
matching the lattice results accurately with the continuum, the provision of 
the same quantity in perturbation theory to as high a loop order as is
computationally possible is important. Indeed the RI${}^\prime$/SMOM lends
itself naturally to this problem of measuring forward matrix elements and hence
generalized parton distribution functions. For instance, there has been
significant lattice work in this general area represented by
\cite{13,14,15,16,17,18,19,20,21,22,23,24,25,26,27,28,29}. As the second moment
of the flavour non-singlet Wilson operator has been treated in \cite{9}, it is 
the purpose of this article to extend that work to the case of the third moment
building on the one loop result of \cite{10}. This is a larger computation as 
the extra covariant derivative means that there is mixing into {\em two} total 
derivative operators, \cite{12}. However, as discussed there these operators 
are actually related to the total derivative of both second moment operators. 
In practice this means that the anomalous dimensions should be the same but 
only some of the amplitudes of the Green's functions will be related. This 
latter property is due to the Lorentz index imbalance between operators of
different moments. In addition to focusing on the RI${}^\prime$/SMOM scheme
structure of the Green's function we will also provide the full set of $\MSbar$
results at the symmetric subtraction point. Aside from being the reference 
renormalization scheme we record these because some lattice groups choose to 
devise their own RI${}^\prime$/SMOM type scheme and then convert their results 
to the $\MSbar$ scheme. Therefore, they would perform the high energy matching 
to perturbation theory in the continuum in $\MSbar$. Whilst all the lattice 
computations are performed in the Landau gauge, our results will be in an 
arbitrary linear covariant gauge. This is because the presence of the gauge 
parameter is partly used as a checking procedure within the symbolic 
manipulation programmes we have used. In addition as was noted in \cite{9,10} 
when one has two or more free Lorentz indices on the inserted operator, then 
there is not a unique way of defining RI${}^\prime$/SMOM. This is mainly to do 
with what combination of basis tensors one uses to write the Green's function 
itself in. Different basis choices could lead to different RI${}^\prime$/SMOM 
schemes. Indeed it might be possible to choose one in such a way that the 
associated conversion functions have a better convergence than the analogous 
RI${}^\prime$ one. One final point concerning the third moment of the Wilson 
operator and that is that the vector current is present in the set of operators
at this level through one of the total derivative operators, \cite{12}. As this
current is physical then its renormalization is determined by general 
properties of renormalization theory and the Slavnov-Taylor identities. 
Therefore, we will pay special attention to that to ensure there is 
consistency.

The article is organized as follows. We devote the next section to the 
background to the set of operators we will renormalize and the structure of the
Green's function we will evaluate to two loops in both schemes as well as a
summary of the technical machinery we employed to perform the computation. The 
details of the two loop renormalization as well as the RI${}^\prime$/SMOM 
anomalous dimension mixing matrix are given in section three. The explicit 
forms of all the two loop conversion functions are given in the subsequent 
section. These are then used in section five to determine the Landau gauge 
anomalous dimensions for the diagonal and some off-diagonal entries of the 
mixing matrix at {\em three} loops. The expressions for the amplitudes of the
Green's function we compute are given in section six for both the $\MSbar$ and
RI${}^\prime$/SMOM schemes with conclusions provided in section seven. An
appendix gives the explicit form of the tensor basis used for the Green's
function as well as the projection matrix used to extract the amplitudes. 

\sect{Background.}

We begin by recalling the formalism and properties of the operators which we
will consider here. The notation we use is the same as in previous articles,
\cite{9,10,11,12}, and we will denote our three operators with the shorthand 
notation,
\begin{equation}
W_3 ~ \equiv ~ {\cal S} \bar{\psi} \gamma^\mu D^\nu D^\sigma \psi ~~,~~ 
\partial W_3 ~ \equiv ~ {\cal S} \partial^\mu \left( \bar{\psi} \gamma^\nu
D^\sigma \psi \right) ~~,~~
\partial \partial W_3 ~ \equiv ~ {\cal S} \partial^\mu \partial^\nu \left(
\bar{\psi} \gamma^\sigma \psi \right)
\label{w3ops}
\end{equation}
where all derivatives act to the right and ${\cal S}$ denotes both
symmetrization and tracelessness, in $d$ spacetime dimensions, in all Lorentz 
indices. In particular 
\begin{eqnarray}
{\cal S} {\cal O}^i_{\mu\nu\sigma} &=&
{\cal O}^i_{S~\mu\nu\sigma} ~-~ \frac{1}{(d+2)} \left[
\eta_{\mu\nu} {\cal O}^{i~~~\rho}_{S~\sigma~\rho} ~+~
\eta_{\nu\sigma} {\cal O}^{i~~~\rho}_{S~\mu~\rho} ~+~
\eta_{\sigma\mu} {\cal O}^{i~~~\rho}_{S~\nu~\rho} \right]
\end{eqnarray}
with
\begin{eqnarray}
{\cal O}^i_{S~\mu\nu\sigma} &=& \frac{1}{6} \left[ 
{\cal O}^i_{\mu\nu\sigma} ~+~ {\cal O}^i_{\nu\sigma\mu} ~+~
{\cal O}^i_{\sigma\mu\nu} ~+~ {\cal O}^i_{\mu\sigma\nu} ~+~
{\cal O}^i_{\sigma\nu\mu} ~+~ {\cal O}^i_{\nu\mu\sigma} \right]
\end{eqnarray}
where the basic operators, ${\cal O}^i_{\mu\nu\sigma}$, are 
\begin{equation} 
{\cal O}^{W_3}_{\mu\nu\sigma} ~=~ \bar{\psi} \gamma_\mu D_\nu D_\sigma 
\psi ~~,~~ 
{\cal O}^{\partial W_3}_{\mu\nu\sigma} ~=~ \partial_\mu \left( \bar{\psi} 
\gamma_\nu D_\sigma \psi \right) ~~,~~
{\cal O}^{\partial \partial W_3}_{\mu\nu\sigma} ~=~ \partial_\mu \partial_\nu 
\left( \bar{\psi} \gamma_\sigma \psi \right) ~.
\end{equation}
As in previous articles the symbol $W_3$ at times will be referred to as the 
level and in those instances, which will be clear from the context, indicates 
the set of the three operators. For example, it appears here in the 
superscript. The specific choice of level $W_3$ operators is dictated by the 
need to have three independent operators. One could have excluded one of the 
total derivative operators in favour of one with the covariant derivative 
acting on the anti-quark field but we retain consistency with lower moment 
operators here, \cite{9,10,12}. In (\ref{w3ops}) we have suppressed the flavour 
indices and emphasise that we are using flavour non-singlet operators. If the 
operators were flavour singlet then one would require gluonic and ghost 
operators to have a closed set under renormalization. In the flavour 
non-singlet case there is mixing but only within the set (\ref{w3ops}) which is
also closed under renormalization. However, the mixing is only relevant when 
the parent operator of $W_3$ is inserted in a Green's function with a momentum 
flowing through the operator itself, \cite{12}. With the way we have chosen the
set of operators then if no momentum flows through the inserted operator there 
is no contribution to the Green's function from the total derivative operators.
For the RI${}^\prime$/SMOM scheme renormalization, \cite{1}, where there are 
two external momentum, $p$ and $q$, flowing in through the quark legs then 
there is a net momentum flow of $(p+q)$ out of the operator, as illustrated in 
Figure $1$ where the form of the Green's function, $\left\langle \psi(p) 
{\cal O}^i_{\mu\nu\sigma} (-p-q) \bar{\psi}(q) \right\rangle$, we consider is 
given. Since we work at the symmetric subtraction point, \cite{1,2,3}, then  
\begin{equation}
p^2 ~=~ q^2 ~=~ ( p + q )^2 ~=~ -~ \mu^2
\end{equation}
which imply
\begin{equation}
pq ~=~ \frac{1}{2} \mu^2
\end{equation}
where $\mu$ is the common mass scale which in this case is equivalent to the 
mass scale introduced to ensure that the coupling constant is dimensionless in
dimensional regularization which we use throughout.

\begin{figure}[ht]
\hspace{6cm}
\epsfig{file=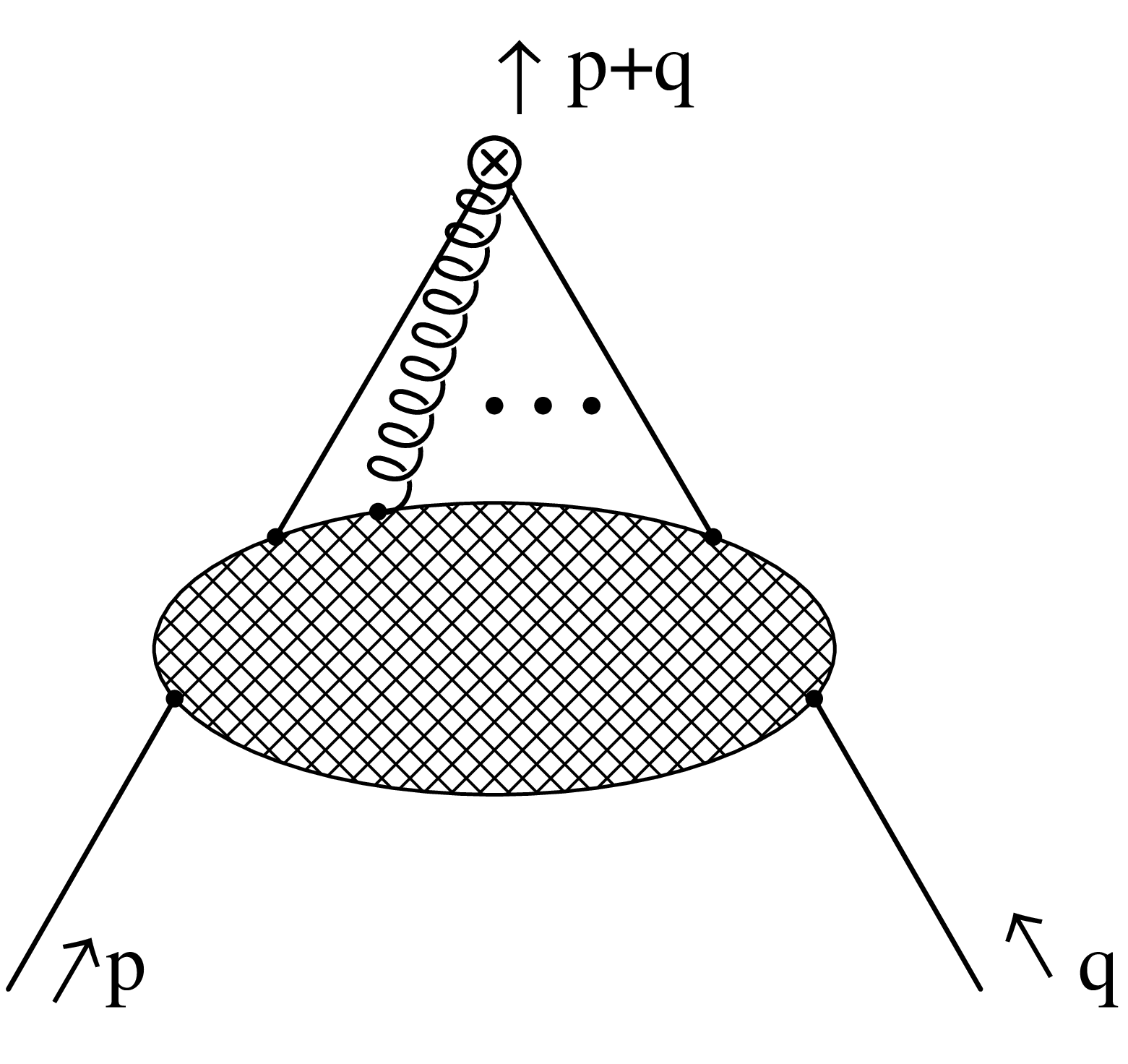,height=5cm}
\vspace{0.5cm}
\caption{The Green's function $\left\langle \psi(p) {\cal O}^i_{\mu\nu\sigma}
(-p-q) \bar{\psi}(q) \right\rangle$.}
\end{figure}

The renormalization of the operators of (\ref{w3ops}) is accommodated by the
mixing matrix of renormalization constants defined by  
\begin{equation}
Z^{W_3}_{ij} ~=~ \left(
\begin{array}{ccc}
Z^{W_3}_{11} & Z^{W_3}_{12} & Z^{W_3}_{13} \\
0 & Z^{W_3}_{22} & Z^{W_3}_{23} \\
0 & 0 & Z^{W_3}_{33} \\
\end{array}
\right) 
\end{equation}
where the subscripts label the operators in the order $W_3$, $\partial W_3$ and
$\partial \partial W_3$. The upper triangularity of the matrix follows from the
choice of our operators in (\ref{w3ops}) and simplifies the computation when 
compared to the extraction of all nine elements for another choice. Associated 
with the renormalization constants are the anomalous dimensions which are 
encoded in a parallel matrix with the formal definition  
\begin{equation}
\gamma^{W_3}_{ij}(a,\alpha) ~=~ -~ \mu \frac{d ~}{d \mu} \ln Z^{W_3}_{ij}
\label{matdef}
\end{equation}
with
\begin{equation}
\mu \frac{d~}{d\mu} ~=~ \beta(a) \frac{\partial ~}{\partial a} ~+~
\alpha \gamma_\alpha(a,\alpha) \frac{\partial ~}{\partial \alpha} ~.
\end{equation}
Here $a$ is the shorthand for the coupling constant via $a$~$=$~$g^2/(16\pi^2)$
and $\alpha$ is the canonical gauge parameter associated with a linear 
covariant gauge fixing with $\alpha$~$=$~$0$ corresponding to the Landau gauge.
Further, $\beta(a)$ is the $\beta$-function and $\gamma_\alpha(a,\alpha)$ is 
the anomalous dimension of the gauge parameter where we retain the conventions 
of \cite{8}. Whilst the anomalous dimensions for gauge invariant operators are
independent of the gauge parameter in massless renormalization schemes, such as
$\MSbar$, we have retained it here since the RI${}^\prime$/SMOM scheme is a 
mass dependent scheme and therefore the anomalous dimensions will depend on the
choice of gauge as will be evident later and hence is indicated in 
(\ref{matdef}). The explicit values for the $\MSbar$ anomalous dimensions have 
already been determined in the $\MSbar$ scheme and we record them as they are 
required for constructing the RI${}^\prime$/SMOM three loop anomalous 
dimensions later. From \cite{12}
\begin{eqnarray}
\left. \gamma^{W_3}_{11}(a) \right|_{\MSbars} &=& \frac{25}{6} C_F a ~+~ 
\frac{1}{432} \left[ 8560 C_A C_F - 2035 C_F^2 - 3320 C_F T_F \Nf \right] a^2 
\nonumber \\
&& +~ \frac{1}{15552} \left[ \left( 285120 \zeta(3) + 1778866 \right) C_A^2 C_F
- \left( 855360 \zeta(3) + 311213 \right) C_A C_F^2 \right. \nonumber \\
&& \left. ~~~~~~~~~~~~~-~ \left( 1036800 \zeta(3) + 497992 \right)
C_A C_F T_F \Nf \right. \nonumber \\
&& \left. ~~~~~~~~~~~~~+~ \left( 570240 \zeta(3) - 244505 \right) C_F^3
+ \left( 1036800 \zeta(3) - 814508 \right) C_F^2 T_F \Nf \right. \nonumber \\
&& \left. ~~~~~~~~~~~~~-~ 82208 C_F T_F^2 \Nf^2 \right] a^3 ~+~ O(a^4) ~,  
\nonumber \\
\left. \gamma^{W_3}_{12}(a) \right|_{\MSbars} &=& -~ \frac{3}{2} C_F a ~+~ 
\frac{1}{144} \left[ 81 C_F^2 - 848 C_A C_F + 424 C_F T_F \Nf \right] a^2 ~+~ 
O(a^3) ~, \nonumber \\
\left. \gamma^{W_3}_{13}(a) \right|_{\MSbars} &=& -~ \frac{1}{2} C_F a ~+~ 
\frac{1}{144} \left[ 103 C_F^2 - 388 C_A C_F + 104 C_F T_F \Nf \right] a^2 ~+~ 
O(a^3) ~, \nonumber \\
\left. \gamma^{W_3}_{22}(a) \right|_{\MSbars} &=& \frac{8}{3} C_F a ~+~ 
\frac{1}{27} \left[ 376 C_A C_F - 112 C_F^2 - 128 C_F T_F \Nf \right] a^2 
\nonumber \\
&& +~ \frac{1}{243} \left[ \left( 5184 \zeta(3) + 20920 \right) C_A^2 C_F
- \left( 15552 \zeta(3) + 8528 \right) C_A C_F^2 \right. \nonumber \\
&& \left. ~~~~~~~~~~-~ \left( 10368 \zeta(3) + 6256 \right) C_A C_F T_F \Nf
+ \left( 10368 \zeta(3) - 560 \right) C_F^3 \right. \nonumber \\
&& \left. ~~~~~~~~~~+~ \left( 10368 \zeta(3) - 6824 \right) C_F^2 T_F \Nf
- 896 C_F T_F^2 \Nf^2 \right] a^3 ~+~ O(a^4) ~, \nonumber \\
\left. \gamma^{W_3}_{23}(a) \right|_{\MSbars} &=& -~ \frac{4}{3} C_F a ~+~ 
\frac{1}{27} \left[ 56 C_F^2 - 188 C_A C_F + 64 C_F T_F \Nf \right] a^2 
\nonumber \\
&& +~ \frac{1}{243} \left[ \left( 7776 \zeta(3) + 4264 \right) C_A C_F^2
- \left( 2592 \zeta(3) + 10460 \right) C_A^2 C_F \right. \nonumber \\
&& \left. ~~~~~~~~~~+~ \left( 5184 \zeta(3) + 3128 \right) C_A C_F T_F \Nf
- \left( 5184 \zeta(3) - 280 \right) C_F^3 \right. \nonumber \\
&& \left. ~~~~~~~~~~-~ \left( 5184 \zeta(3) - 3412 \right) C_F^2 T_F \Nf
+ 448 C_F T_F^2 \Nf^2 \right] a^3 ~+~ O(a^4) ~, \nonumber \\
\left. \gamma^{W_3}_{33}(a) \right|_{\MSbars} &=& O(a^4)
\label{anom3}
\end{eqnarray}
where $\zeta(z)$ is the Riemann zeta function and the colour group Casimirs are
defined as 
\begin{equation}
T^a T^a ~=~ C_F ~~,~~ f^{acd} f^{bcd} ~=~ C_A \delta^{ab} ~~,~~ 
\mbox{Tr} \left( T^a T^b \right) ~=~ \delta^{ab} T_F
\end{equation}
where $1$~$\leq$~$a$~$\leq$~$\NA$, $\NA$ is the dimension of the adjoint
representation and $\Nf$ is the number of massless quarks. However, only the
diagonal elements and the $23$ elements are known to three loops. The 
off-diagonal elements for the first row were only determined to two loops. The
reason for this is that to extract the mixing matrix renormalization
constants, \cite{12}, one has to consider various momentum routings of the
operator inserted in the Green's function of Figure $1$ as well as exploit the
fact that there must be no terms such as $(\ln(p^2/\mu^2))/\epsilon$, where 
$d$~$=$~$4$~$-$~$2\epsilon$ and $p$ is the sole momentum flowing through 
external legs. This latter criterion establishes relations between the 
off-diagonal counterterms. For the $n$~$=$~$2$ moment operator one had 
sufficient linear equations in order to determine the three loop mixing matrix,
\cite{12}. With the increase in moment one requires a further linear relation 
in order to resolve the counterterms at three loops. This can only be achieved 
by a {\em four} loop renormalization for $W_3$ which was not considered in 
\cite{12}. 

Throughout we will use the convention that the results will be expressed in the
RI${}^\prime$/SMOM scheme, using RI${}^\prime$/SMOM variables, unless otherwise 
specified as shown in (\ref{anom3}). As we will be mapping between the 
$\MSbar$ and RI${}^\prime$/SMOM schemes we recall that the variables are not
necessarily the same in each scheme. However, the relations are known to three
loops, \cite{8}, and are  
\begin{equation}
a_{\mbox{\footnotesize{RI$^\prime$}}} ~=~
a_{\mbox{\footnotesize{$\MSbars$}}} ~+~ O \left(
a_{\mbox{\footnotesize{$\MSbars$}}}^5 \right)
\label{aconv}
\end{equation}
and 
\begin{eqnarray}
\alpha_{\mbox{\footnotesize{RI$^\prime$}}}
&=& \left[ 1 + \left( \left( - 9 \alpha_{\mbox{\footnotesize{$\MSbars$}}}^2
- 18 \alpha_{\mbox{\footnotesize{$\MSbars$}}} - 97 \right) C_A + 80 T_F \Nf
\right) \frac{a_{\mbox{\footnotesize{$\MSbars$}}}}{36} \right. \nonumber \\
&& \left. ~+~ \left( \left( 18 \alpha_{\mbox{\footnotesize{$\MSbars$}}}^4
- 18 \alpha_{\mbox{\footnotesize{$\MSbars$}}}^3
+ 190 \alpha_{\mbox{\footnotesize{$\MSbars$}}}^2
- 576 \zeta(3) \alpha_{\mbox{\footnotesize{$\MSbars$}}}
+ 463 \alpha_{\mbox{\footnotesize{$\MSbars$}}} + 864 \zeta(3) - 7143 \right)
C_A^2 \right. \right. \nonumber \\
&& \left. \left. ~~~~~~~+~ \left( -~ 320
\alpha_{\mbox{\footnotesize{$\MSbars$}}}^2
- 320 \alpha_{\mbox{\footnotesize{$\MSbars$}}} + 2304 \zeta(3)
+ 4248 \right) C_A T_F \Nf \right. \right. \nonumber \\
&& \left. \left. ~~~~~~~+~ \left( -~ 4608 \zeta(3) + 5280 \right) C_F T_F \Nf
\right) \frac{a^2_{\mbox{\footnotesize{$\MSbars$}}}}{288} \right]
\alpha_{\mbox{\footnotesize{$\MSbars$}}} ~+~ O \left(
a^3_{\mbox{\footnotesize{$\MSbars$}}} \right) 
\label{alconv}
\end{eqnarray}
which implies that the Landau gauge is preserved under mappings between 
schemes. These relations derive from \cite{8} where the full three loop 
renormalization of QCD in a linear covariant gauge in the RI${}^\prime$ scheme
was recorded. That scheme was defined by ensuring that all the $2$-point 
function renormalizations were performed in such a way that there were no 
$O(a)$ corrections at the subtraction point. So included within this is the 
gluon $2$-point function which determines the renormalization of $\alpha$. The
coupling constant renormalization was performed using $3$-point function 
renormalization but those functions and higher are renormalized in the $\MSbar$
fashion whence the trivial mapping of (\ref{aconv}). The RI${}^\prime$/SMOM 
scheme differs from RI${}^\prime$ in that the $3$-point and higher functions 
are rendered finite by ensuring there are no $O(a)$ corrections at the 
symmetric subtraction point. We should note that in \cite{2,3} the 
renormalization of $\alpha$ was assumed to have been carried out in the 
$\MSbar$ scheme. Whilst this means that Landau gauge expressions will be 
similar, if one was comparing the $\alpha$ dependent parts for the same 
quantities in the definition of \cite{1,2,3} and that used here, then they 
would not be in agreement for non-zero $\alpha$. 

The Green's function we will compute with the level $W_3$ operators inserted is
illustrated in Figure $1$ where there are three Lorentz indices associated with
the operator insertion which is denoted by the circle containing the cross.
Therefore, at the symmetric subtraction point the Green's function has to be
decomposed into a basis of Lorentz tensors which obey the same tracelessness 
and total symmetry as the original operator. The choice of basis tensors was
given in \cite{10} for the one loop computation and we retain the same basis 
here. More explicitly at the subtraction point we formally write
\begin{equation}
\left. \left\langle \psi(p) {\cal O}^i_{\mu \nu \sigma}(-p-q)
\bar{\psi}(q) \right\rangle \right|_{p^2 = q^2 = - \mu^2} ~=~ \sum_{k=1}^{14}
{\cal P}^i_{(k) \, \mu \nu \sigma }(p,q) \,
\Sigma^{{\cal O}^i}_{(k)}(p,q) 
\end{equation}
where ${\cal P}^i_{(k) \, \mu \nu \sigma }(p,q)$ are the basis tensors. For
$W_3$ there are fourteen of these and as they are cumbersome we have provided
the explicit forms in the Appendix. It should be noted that the tensor basis we
use is defined at the symmetric subtraction point and is the same basis for 
each of the operators of level $W_3$. When the squares of the external momenta 
of the Green's function are not all equal then the basis of tensors will be 
much larger. The objects which we will concentrate on computing are the 
amplitudes associated with each tensor for each $W_3$ operator insertion. These
are $\Sigma^{{\cal O}^i}_{(k)}(p,q)$ and defined by the projection method,
\cite{10},  
\begin{equation}
\Sigma^{{\cal O}^i}_{(k)}(p,q) ~=~ {\cal M}^i_{kl}
{\cal P}^{i ~\, \mu \nu \sigma}_{(l)}(p,q) \left. \left\langle \psi(p) 
{\cal O}^i_{\mu \nu \sigma}(-p-q) \bar{\psi}(q) \right\rangle 
\right|_{p^2=q^2=-\mu^2} ~.
\end{equation}
Here ${\cal M}^i_{kl}$ is a dimension fourteen matrix of rational polynomials
in $d$. It is computed by inverting the matrix ${\cal N}^i_{kl}$ which is 
defined by
\begin{equation}
{\cal N}^i_{kl} ~=~ \left. {\cal P}^i_{(k) \, \mu \nu \sigma}(p,q)
{\cal P}^{i ~\, \mu \nu \sigma}_{(l)}(p,q) \right|_{p^2=q^2=-\mu^2} ~.
\end{equation}
Due to the large number of basis tensors this matrix is equally cumbersome and
is also recorded explicitly in Appendix A.

In the choice of basis tensors we used the generalized $\gamma$-matrices 
$\Gamma_{(m)}^{\mu_1 \ldots \mu_m}$, \cite{30,31,32,33,34}, which are defined 
by 
\begin{equation}
\Gamma_{(n)}^{\mu_1 \ldots \mu_n} ~=~ \gamma^{[\mu_1} \ldots \gamma^{\mu_n]}
\end{equation}
where the notation includes an overall factor $1/n!$. These are totally
antisymmetric and form the generalization of the $\gamma$-matrices in 
$d$-dimensional spacetime. Athough we are ultimately interested in four 
dimensions they are a more appropriate object to use to build the tensor basis 
as the generalized $\gamma$-space naturally partitions into regions defined by
the number of free Lorentz indices. The justification for this lies in the
observation, \cite{30,31,32}, 
\begin{equation}
\mbox{tr} \left( \Gamma_{(m)}^{\mu_1 \ldots \mu_m}
\Gamma_{(n)}^{\nu_1 \ldots \nu_n} \right) ~ \propto ~ \delta_{mn}
I^{\mu_1 \ldots \mu_m \nu_1 \ldots \nu_n}
\label{gamtr}
\end{equation}
where $I^{\mu_1 \ldots \mu_m \nu_1 \ldots \nu_n}$ is the unit matrix in this
$\Gamma$-space. Indeed this is why (\ref{matw3}) has a $\Gamma_{(1)}^\mu$ and
$\Gamma_{(3)}^{\mu\nu\sigma}$ subspace. Though we will always use $\gamma^\mu$ 
as the object rather than the generalized object. Hence, when constructing the
tensor basis one can be confident that a complete set has been used. For
instance, the antisymmetry means that at most two external momenta can be
contracted with $\Gamma_{(n)}^{\mu_1 \ldots \mu_n}$ for $n$~$\geq$~$2$. So the
Lorentz aspect of the basis is built from combinations of tensors from the set
$\{ \eta^{\mu\nu}, p^\mu, q^\mu, \Gamma_{(n)}^{\mu_1 \ldots \mu_n} \}$. The
choice of the fourteen given in Appendix A were also derived from the explicit
one loop computation, \cite{10}. In other words the one loop Feynman diagrams 
were expressed in terms of the basis Lorentz tensor Feynman integrals which
were evaluated directly. The expressions for these were substituted back into 
the original computation and the Lorentz indices contracted producing the 
fourteen basis tensors, \cite{10}. Therefore, given that we use the basis here 
but do not construct all possible two loop basis Lorentz tensor master 
integrals. Instead we use the projection method. 

The machinery used to perform the computations are the symbolic manipulation
language {\sc Form}, \cite{35}. The $3$ one loop and $37$ two loop Feynman
diagrams are generated in electronic format using the {\sc Qgraf} package,
\cite{36}, and then Lorentz and colour group indices appended. As the Green's
function contains two external momenta then packages such as {\sc Mincer}, 
\cite{37}, are not applicable. Instead we first rewrite all the diagrams as 
scalar integrals where the numerator tensor structure is rewritten in terms of 
the denominator propagators in as far as this is possble. In this form we then
used the Laporta algorithm, \cite{38}. This creates a set of linear relations 
between all integrals with a certain specification for each appropriate 
momentum topology. The relations are established by integration by parts and 
Lorentz identities and the integrals of interest are rewritten in terms of a 
relatively small set of basic scalar master Feynman integrals. For the two loop
Green's function these have already been established to the order in $\epsilon$
we need in order to find the finite part. This is not an empty statement
because in the huge set of linear equations spurious poles in $\epsilon$ can 
emerge in the factor multiplying the master integrals. To handle such an 
enormous set of equations requires computer technology. Several packages are 
available but we used the {\sc Reduze} programme, \cite{39}, based on 
{\sc Ginac}, \cite{40}, which is written in C$++$. It turns out that for the 
$37$ two loop diagrams there are only two basic momentum topologies. In other 
words the momentum routing of all the diagrams could be expressed in terms of 
one or other of these configurations. Also given that we had constructed the 
algorithm for the lower moment operators, $W_2$, it was straightforward to lift
the extra integral relations needed for $W_3$ out of the earlier database 
created for $W_2$. Therefore, we are reasonably confident that our procedure is
consistent as the same routines were used for the quark current operator 
renormalization which agreed with the two loop results of \cite{2,3} for the 
scalar and tensor operators.  

\sect{Renormalization.}

The renormalization of the $W_3$ operators has been outlined in \cite{12} and
we briefly recall the procedure. First, we compute the Green's function with 
all the parameters being regarded as bare. This is the simplest way to proceed 
when dealing with automatic symbolic manipulation computer programmes. The
renormalized variables are then introduced via the associated renormalization
constant in the canonical way, \cite{41}. This ensures that the counterterms 
are correctly embedded within the Feynman diagrams automatically. For the 
operators themselves the renormalized versions are also introduced by a similar
rescaling involving their renormalization constant. However, unlike the fields 
and parameters this renormalization is not multiplicative due to the mixing,
\cite{12}. Therefore, when the renormalized operators are introduced we include
the operators into which it mixes. Consequently poles in $\epsilon$ will appear
in various channels of the tensor basis but not all. They have to be removed by
the criterion of the scheme in question. For $\MSbar$ this is achieved in the
standard way. However, for the RI${}^\prime$/SMOM scheme it is more involved.
If, for the moment we concentrate on a generic tensor channel which contains
divergences, and denote that channel by $0$, then the renormalization 
condition is, \cite{1},   
\begin{equation}
\left. \lim_{\epsilon\rightarrow 0} \left[
Z^{\mbox{\footnotesize{RI$^\prime$}}}_\psi
Z^{\mbox{\footnotesize{RI$^\prime$/SMOM}}}_{\cal O} \Sigma^{\cal O}_{(0)}(p,q)
\right] \right|_{p^2 \, = \, q^2 \, = \, - \mu^2} ~=~ 1 ~.
\label{rencon}
\end{equation}
The quark wave function renormalization is performed in the RI${}^\prime$
scheme but as we are now dealing with a $3$-point function then the operator
renormalization constant obeys the same condition. The ethos for the 
regularization invariant schemes is that after renormalization there are no 
$O(a)$ corrections at the subtraction point which here is the symmetric point,
\cite{1}. So both the quark $2$-point function and the Green's function we are
considering have no $O(a)$ corrections. For $W_3$ we need to qualify 
(\ref{rencon}) by saying that the generic renormalization constant could
represent a combination of counterterms which appear in the mixing matrix.
Therefore to determine all the information to compute the anomalous dimension
mixing matrix requires the solution of a set of linear equations. This was also
a feature of the original $\MSbar$ renormalization of the operators in 
\cite{12}. Whilst (\ref{rencon}) is the condition which determines the operator
renormalization constant to avoid confusion it is important to note that the 
{\em full} Green's function is multiplied by the mixing matrix of 
renormalization constants. So after the renormalization constants have been 
fixed the finite parts of all the amplitudes are affected by the 
renormalization.

Having outlined the general aspects of the renormalization we have to discuss
the specific renormalization of $\partial \partial W_3$ which requires special
attention. This is because that operator is related to the divergence of the
vector current. The vector current is a special operator in that it is 
physical. Therefore, not only does it not get renormalized to all orders in
perturbation theory but the renormalization constant is the {\em same} in all
schemes. See, for example, \cite{42} for a summary of this. This is a 
consequence of gauge symmetry and effected by the Slavnov-Taylor identities. 
Therefore, our RI${}^\prime$ scheme renormalization of $\partial \partial W_3$ 
must be consistent with these general principles. So from the definition of 
$\partial \partial W_3$ we have to project out that piece which corresponds to 
the total derivative of the divergence of the vector current. Contracting the 
Green's function containing the $\partial \partial W_3$ operator with 
$(p+q)_\mu \eta_{\nu\sigma}$ does not produce anything non-trivial due to the 
traceless condition. Instead we contract with $(p+q)_\mu (p+q)_\nu 
(p+q)_\sigma$ and this combination of vectors is chosen since that is the 
momentum flow through the operator itself. This produces a non-trivial 
combination of the amplitudes. Ordinarily this would produce an expression 
where there are poles in $\epsilon$. Indeed the individual amplitudes have 
divergences. However, as the underlying operator is finite there are no 
divergences in $\epsilon$. Whilst this is consistent with earlier remarks one 
has also to ensure that the Slavnov-Taylor identity is actually satisfied and 
this relates to the finite part of this projection. Computing in the $\MSbar$ 
scheme, where the scheme dependence is in effect the quark wave function
renormalization from (\ref{rencon}), we find that the combination of amplitudes 
of the contraction proportional to $\pslash$ is, \cite{10},  
\begin{eqnarray}
&& \frac{3[d-6]}{4[d+2]} \left. \Sigma^{\partial\partial W_3}_{(1)}(p,q) 
\right|_{\MSbars} ~+~ 
\frac{3}{2} \left. \Sigma^{\partial\partial W_3}_{(2)}(p,q) 
\right|_{\MSbars} ~+~
\frac{3[d-4]}{4[d+2]} \left. \Sigma^{\partial\partial W_3}_{(3)}(p,q) 
\right|_{\MSbars} \nonumber \\
&& -~ \frac{[d-10]}{8[d+2]} \left. \Sigma^{\partial\partial W_3}_{(4)}(p,q) 
\right|_{\MSbars} ~-~ 
\frac{3}{8} \left. \Sigma^{\partial\partial W_3}_{(5)}(p,q) 
\right|_{\MSbars} ~-~
\frac{3}{8} \left. \Sigma^{\partial\partial W_3}_{(6)}(p,q) 
\right|_{\MSbars} \nonumber \\
&& -~ \frac{[d-10]}{8[d+2]} \left. \Sigma^{\partial\partial W_3}_{(7)}(p,q) 
\right|_{\MSbars} \nonumber \\ 
&& =~ -~ \frac{1}{2} ~-~ \frac{1}{2} \alpha C_F a \nonumber \\
&& ~~~~+~ \left[ \left[ \frac{3}{2} \zeta(3) - \frac{41}{8} + \frac{3}{2} 
\zeta(3) \alpha - \frac{13}{4} \alpha - \frac{9}{16} \alpha^2 \right] 
C_F C_A ~+~ \frac{7}{4} C_F T_F \Nf ~+~ \frac{5}{16} C_F^2 \right] a^2
\nonumber \\
&& ~~~~+~ O(a^3) ~.
\end{eqnarray}
A similar expression for the piece involving $\qslash$ produces the same
outcome. Clearly it is proportional to the finite part of the quark $2$-point 
function after renormalization in the $\MSbar$ scheme. Indeed the equality of 
the $\pslash$ and $\qslash$ parts with the unit renormalization constant 
required for $\partial \partial W_3$, due to the absence of poles in 
$\epsilon$, as well as the explicit values for the $\MSbar$ amplitudes means 
that our renormalization is consistent with the Slavnov-Taylor identities. 
Repeating the process for the RI${}^\prime$/SMOM scheme with the RI${}^\prime$ 
scheme quark wave function renormalization constant produces a similar result 
which is 
\begin{eqnarray}
&& \frac{3[d-6]}{4[d+2]} \Sigma^{\partial\partial W_3}_{(1)}(p,q) ~+~
\frac{3}{2} \Sigma^{\partial\partial W_3}_{(2)}(p,q) ~+~
\frac{3[d-4]}{4[d+2]} \Sigma^{\partial\partial W_3}_{(3)}(p,q) ~-~
\frac{[d-10]}{8[d+2]} \Sigma^{\partial\partial W_3}_{(4)}(p,q) \nonumber \\
&& -~ \frac{3}{8} \Sigma^{\partial\partial W_3}_{(5)}(p,q) ~-~
\frac{3}{8} \Sigma^{\partial\partial W_3}_{(6)}(p,q) ~-~
\frac{[d-10]}{8[d+2]} \Sigma^{\partial\partial W_3}_{(7)}(p,q) ~=~
-~ \frac{1}{2} ~+~ O(a^3) ~.
\end{eqnarray}
Clearly there are no $O(a)$ corrections which is in agreement with the 
RI${}^\prime$ scheme quark $2$-point function after renormalization. The fact
that when this combination is computed that there were no poles in $\epsilon$
means the renormalization of $\partial \partial W_3$ also has a unit
renormalization constant in this scheme consistent with the physicality of the
operator itself.   

The procedure to determine the remaining renormalization constants of the
mixing matrix is to identify a set of amplitudes or their combinations to which
one can apply the condition (\ref{rencon}). As noted in \cite{9,10} for the
RI${}^\prime$/SMOM scheme there is not a unique way of doing this. One could
use the same combinations for $W_3$ and $\partial W_3$ as $\partial \partial 
W_3$. However, as there are five renormalization constants to determine one
would require four other conditions. Instead we extend to two loops the 
RI${}^\prime$/SMOM scheme definition given in \cite{10}. Here the first three
channels are rendered finite by the generic condition (\ref{rencon}). In 
addition to the lack of uniqueness in defining the scheme because of the choice
of amplitudes there is also the issue that our basis choice of tensors defining
the amplitude channels is not unique. One could have defined a different set of
basis tensors, though with the same symmetries as the operators themselves.
Then one would have another definition of RI${}^\prime$/SMOM. The choice we
make here is in some sense a minimal choice and one could readily make other 
more exotic ones. However, following this particular path we have determined 
the renormalization constants and have encoded them in the mixing matrix of 
anomalous dimensions. We have 
\begin{eqnarray}
\left. \frac{}{} \gamma^{W_3}_{11}(a,\alpha)
\right|_{\mbox{\footnotesize{RI$^\prime$/SMOM}}} &=& \frac{25}{6} C_F a
\nonumber \\
&& +~ \left[ \left[ ( 384 \alpha^2 + 1152 \alpha - 3168 ) \psi^\prime(\third)
- ( 256 \alpha^2 + 786 \alpha - 2112 ) \pi^2 \right. \right. \nonumber \\
&& \left. \left. ~~~~~-~ 288 \alpha^2 - 864 \alpha + 66204 \right] C_A ~-~
6105 C_F \right. \nonumber \\
&& \left. ~~~~+ \left[ 1152 \psi^\prime(\third) - 768 \pi^2 - 24696 \right]
T_F \Nf \right] \frac{C_F a^2}{1296} ~+~ O(a^3) ~, \nonumber \\
\left. \frac{}{} \gamma^{W_3}_{12}(a,\alpha)
\right|_{\mbox{\footnotesize{RI$^\prime$/SMOM}}} &=& -~ \frac{3}{2} C_F a
\nonumber \\
&& +~ \left[ \left[ ( 2112 - 864 \alpha - 288 \alpha^2 ) \psi^\prime(\third)
+ ( 192 \alpha^2 + 576 \alpha - 1408 ) \pi^2 \right. \right. \nonumber \\
&& \left. \left. ~~~~~+~ 216 \alpha^2 + 648 \alpha - 63684 \right] C_A ~+~
2187 C_F \right. \nonumber \\
&& \left. ~~~~+ \left[ 26280 + 512 \pi^2 - 768 \psi^\prime(\third) \right]
T_F \Nf \right] \frac{C_F a^2}{3888} ~+~ O(a^3) ~, \nonumber \\
\left. \frac{}{} \gamma^{W_3}_{13}(a,\alpha)
\right|_{\mbox{\footnotesize{RI$^\prime$/SMOM}}} &=& -~ \frac{1}{2} C_F a
\nonumber \\
&& +~ \left[ \left[ ( 288 \alpha^2 + 864 \alpha - 1056 ) \psi^\prime(\third)
- ( 192 \alpha^2 + 576 \alpha - 704 ) \pi^2 \right. \right. \nonumber \\
&& \left. \left. ~~~~~-~ 1188 \alpha^2 - 3564 \alpha - 10872 \right] C_A 
\right. \nonumber \\
&& \left. ~~~~+ \left[ ( 2592 - 2784 \alpha ) \psi^\prime(\third)
+ ( 1856 \alpha - 1728 ) \pi^2 \right. \right. \nonumber \\
&& \left. \left. ~~~~~+~ 8244 \alpha - 9207 \right] C_F \right. \nonumber \\
&& \left. ~~~~+ \left[ 384 \psi^\prime(\third) - 256 \pi^2 + 2952 \right]
T_F \Nf \right] \frac{C_F a^2}{3888} ~+~ O(a^3) ~, \nonumber \\
\left. \frac{}{} \gamma^{W_3}_{22}(a,\alpha)
\right|_{\mbox{\footnotesize{RI$^\prime$/SMOM}}} &=& \frac{8}{3} C_F a
\nonumber \\
&& +~ \left[ \left[ ( 108 \alpha^2 + 324 \alpha - 924 ) \psi^\prime(\third)
- ( 72 \alpha^2 + 216 \alpha - 616 ) \pi^2 \right. \right. \nonumber \\
&& \left. \left. ~~~~~-~ 81 \alpha^2 - 243 \alpha + 16866 \right] C_A ~-~
2016 C_F \right. \nonumber \\
&& \left. ~~~~+ \left[ 336 \psi^\prime(\third) - 224 \pi^2 - 5976 \right]
T_F \Nf \right] \frac{C_F a^2}{486} ~+~ O(a^3) ~, \nonumber \\
\left. \frac{}{} \gamma^{W_3}_{23}(a,\alpha)
\right|_{\mbox{\footnotesize{RI$^\prime$/SMOM}}} &=& -~ \frac{4}{3} C_F a
\nonumber \\
&& +~ \left[ \left[ 264 \psi^\prime(\third) - 176 \pi^2 - 81 \alpha^2
- 243 \alpha - 6651 \right] C_A \right. \nonumber \\
&& \left. ~~~~+ \left[ 144 \psi^\prime(\third) - 288 \psi^\prime(\third)
\alpha - 288 + 648 \alpha - 96 \pi^2 + 192 \pi^2 \alpha^2 \right] C_F \right.
\nonumber \\
&& \left. ~~~~+ \left[ 64 \pi^2 + 2340 - 96 \psi^\prime(\third) \right]
T_F \Nf \right] \frac{C_F a^2}{486} ~+~ O(a^3) ~, \nonumber \\
\left. \frac{}{} \gamma^{W_3}_{33}(a,\alpha)
\right|_{\mbox{\footnotesize{RI$^\prime$/SMOM}}} &=&  O(a^3) 
\label{andim2}
\end{eqnarray}
where $\psi(z)$ is the derivative of the logarithm of the Euler 
$\Gamma$-function. As a check on our procedures we first verified that the two 
loop $\MSbar$ mixing matrix of anomalous dimensions emerged and agreed with 
\cite{12}. Whilst \cite{12} was a three loop computation it used the 
{\sc Mincer} algorithm, \cite{37}, designed for massless $2$-point function 
renormalization in dimensional regularization. The one loop terms are clearly 
gauge parameter independent but the two and higher loop corrections will depend
on the gauge parameter $\alpha$. This is because in mass dependent 
renormalization schemes, such as RI${}^\prime$/SMOM, the anomalous dimensions 
of gauge invariant operators are not gauge independent in contrast to mass 
independent schemes such as $\MSbar$. A final observation on the calculation is
that the renormalization of $\partial W_3$ is the same as that for $W_2$ which 
was considered in \cite{9}. This is a non-trivial check since one has a 
different set of tensors for the basis due to the imbalance of Lorentz indices 
between the operators. That the same renormalization emerges is partly due to a
similar renormalization scheme choice but also because the set of operators 
$W_n$ are all connected via the tower of operators involving the associated 
total derivative operators. 

\sect{Conversion functions.}

Having constructed the RI${}^\prime$/SMOM mixing matrix of anomalous dimensions
at two loops we can construct the associated conversion functions. These allow
one to translate between renormalization schemes. For us the two schemes will
be RI${}^\prime$/SMOM and $\MSbar$ and general background can be found in, for 
example, \cite{42}. However, as the renormalization of the operators is via a 
mixing matrix one has to extend the theory to allow for this. So the natural 
extension is to a matrix of conversion functions which is formally defined by 
\begin{equation}
C^{W_3}_{ij}(a,\alpha) ~=~
Z^{W_3}_{{ik},\mbox{\footnotesize{RI$^\prime$/SMOM}}}
\left[ Z^{W_3}_{{kj},\mbox{\footnotesize{$\MSbar$}}} \right]^{-1} ~.
\end{equation}
As the parameters $a$ and $\alpha$ are tied to a renormalization scheme we 
have to make a choice and note that in the argument of the conversion functions
the parameters will be in the $\MSbar$ scheme. So in this definition those 
parameters in the RI${}^\prime$/SMOM renormalization constants have to be 
converted to the corresponding $\MSbar$ parameters via (\ref{aconv}) and
(\ref{alconv}). Otherwise aside from being inconsistent one would have a 
non-finite expression due to poles in $\epsilon$. For practical purposes, the
definition translates into  
\begin{eqnarray}
C^{W_3}_{ii}(a,\alpha) &=&
\frac{Z^{W_3}_{ii,\mbox{\footnotesize{RI$^\prime$/SMOM}}}}
{Z^{W_3}_{ii,\mbox{\footnotesize{$\MSbar$}}}} ~~,~~
C^{W_3}_{12}(a,\alpha) ~=~
\frac{Z^{W_3}_{12,\mbox{\footnotesize{RI$^\prime$/SMOM}}}}
{Z^{W_3}_{22,\mbox{\footnotesize{$\MSbar$}}}} ~-~
\frac{Z^{W_3}_{11,\mbox{\footnotesize{RI$^\prime$/SMOM}}}
Z^{W_3}_{12,\mbox{\footnotesize{$\MSbar$}}}}
{Z^{W_3}_{11,\mbox{\footnotesize{$\MSbar$}}}
Z^{W_3}_{22,\mbox{\footnotesize{$\MSbar$}}}} ~, \nonumber \\
C^{W_3}_{13}(a,\alpha) &=&
\frac{Z^{W_3}_{13,\mbox{\footnotesize{RI$^\prime$/SMOM}}}}
{Z^{W_3}_{33,\mbox{\footnotesize{$\MSbar$}}}} ~+~
\frac{Z^{W_3}_{11,\mbox{\footnotesize{RI$^\prime$/SMOM}}} 
Z^{W_3}_{12,\mbox{\footnotesize{$\MSbar$}}}
Z^{W_3}_{23,\mbox{\footnotesize{$\MSbar$}}}}
{Z^{W_3}_{11,\mbox{\footnotesize{$\MSbar$}}}
Z^{W_3}_{22,\mbox{\footnotesize{$\MSbar$}}}
Z^{W_3}_{33,\mbox{\footnotesize{$\MSbar$}}}} \nonumber \\
&& -~ \frac{Z^{W_3}_{11,\mbox{\footnotesize{RI$^\prime$/SMOM}}} 
Z^{W_3}_{13,\mbox{\footnotesize{$\MSbar$}}}}
{Z^{W_3}_{11,\mbox{\footnotesize{$\MSbar$}}}
Z^{W_3}_{33,\mbox{\footnotesize{$\MSbar$}}}} ~-~ 
\frac{Z^{W_3}_{12,\mbox{\footnotesize{RI$^\prime$/SMOM}}} 
Z^{W_3}_{23,\mbox{\footnotesize{$\MSbar$}}}}
{Z^{W_3}_{22,\mbox{\footnotesize{$\MSbar$}}}
Z^{W_3}_{33,\mbox{\footnotesize{$\MSbar$}}}} ~, \nonumber \\
C^{W_3}_{23}(a,\alpha) &=&
\frac{Z^{W_3}_{23,\mbox{\footnotesize{RI$^\prime$/SMOM}}}}
{Z^{W_3}_{33,\mbox{\footnotesize{$\MSbar$}}}} ~-~
\frac{Z^{W_3}_{22,\mbox{\footnotesize{RI$^\prime$/SMOM}}}
Z^{W_3}_{23,\mbox{\footnotesize{$\MSbar$}}}}
{Z^{W_3}_{22,\mbox{\footnotesize{$\MSbar$}}}
Z^{W_3}_{33,\mbox{\footnotesize{$\MSbar$}}}}
\end{eqnarray}
for individual entries of the conversion function matrix where the first term
corresponds to the diagonal and there is no sum over $i$ and $i$~$=$~$1$, $2$ 
or $3$. Equipped with these the explicit values of the matrix elements to two
loops are 
\begin{eqnarray}
C^{W_3}_{11}(a,\alpha) &=& 1 ~+~ \left[ ( 192 \alpha - 216 ) 
\psi^\prime(\third) + ( 144 - 128 \alpha ) \pi^2 - 144 \alpha + 2763 \right] 
\frac{C_F a}{324} \nonumber \\
&& +~ \left[ \left[ 
( 294912 \alpha^2 - 663552 \alpha + 248832 ) (\psi^\prime(\third))^2 
\right. \right. \nonumber \\
&& \left. \left. ~~~~~
+~ ( 884736 \alpha - 393216 \alpha^2 - 331776 ) \psi^\prime(\third) \pi^2
\right. \right. \nonumber \\
&& \left. \left. ~~~~~
+~ ( 13360896 \alpha - 857088 \alpha^2 - 16819488 ) \psi^\prime(\third) 
- ( 5184 + 82944 \alpha ) \psi^{\prime\prime\prime}(\third) 
\right. \right. \nonumber \\
&& \left. \left. ~~~~~
+~ ( 40310784 \alpha - 94058496 ) s_2(\pisix) 
+ ( 188116992 - 80621568 \alpha ) s_2(\pitwo) 
\right. \right. \nonumber \\
&& \left. \left. ~~~~~
+~ ( 156764160 - 67184640 \alpha ) s_3(\pisix) 
+ ( 53747712 \alpha - 125411328 ) s_3(\pitwo) 
\right. \right. \nonumber \\
&& \left. \left. ~~~~~
+~ ( 131072 \alpha^2 - 73728 \alpha + 124416 ) \pi^4 
\right. \right. \nonumber \\
&& \left. \left. ~~~~~
+~ ( 571392 \alpha^2 - 8907264 \alpha + 11212992 ) \pi^2 
\right. \right. \nonumber \\
&& \left. \left. ~~~~~
+~ 616896 \alpha^2 - 3348864 \alpha + 20576997
+ ( 1492992 \alpha - 1679616 ) \Sigma
\right. \right. \nonumber \\
&& \left. \left. ~~~~~
+~ ( 3359232 \alpha + 3919104 ) \zeta(3)
+ ( 279936 \alpha - 653184 ) \frac{\ln^2(3) \pi}{\sqrt{3}}
\right. \right. \nonumber \\
&& \left. \left. ~~~~~
+~ ( 7838208 - 3359232 \alpha ) \frac{\ln(3) \pi}{\sqrt{3}}
+ ( 701568 - 300672 \alpha ) \frac{\pi^3}{\sqrt{3}}
\right] C_F \right. \nonumber \\
&& ~~~~~+ \left. \left[ 
62208 (\psi^\prime(\third))^2 
- 82944 \psi^\prime(\third) \pi^2 
\right. \right. \nonumber \\
&& \left. \left. ~~~~~~~~~
+~ ( 580608 \alpha^2 - 1111968 \alpha + 974592 ) \psi^\prime(\third) 
\right. \right. \nonumber \\
&& \left. \left. ~~~~~~~~~
+~ ( 145152 - 46656 \alpha ) \psi^{\prime\prime\prime}(\third) 
+ ( 65505024 - 25194240 \alpha ) s_2(\pisix) 
\right. \right. \nonumber \\
&& \left. \left. ~~~~~~~~~
+~ ( 50388480 \alpha - 131010048 ) s_2(\pitwo) 
+ ( 41990400 \alpha - 109175040 ) s_3(\pisix) 
\right. \right. \nonumber \\
&& \left. \left. ~~~~~~~~~
+~ ( 87340032 - 33592320 \alpha ) s_3(\pitwo) 
+ ( 124416 \alpha - 359424 ) \pi^4 
\right. \right. \nonumber \\
&& \left. \left. ~~~~~~~~~
-~ ( 387072 \alpha^2 - 741312 \alpha + 649728 ) \pi^2 
\right. \right. \nonumber \\
&& \left. \left. ~~~~~~~~~
-~ 505440 \alpha^2 - 1714608 \alpha + 79566624
\right. \right. \nonumber \\
&& \left. \left. ~~~~~~~~~
+~ ( 746496 \alpha - 1213056 ) \Sigma
+ ( 1679616 \alpha - 24354432 ) \zeta(3)
\right. \right. \nonumber \\
&& \left. \left. ~~~~~~~~~
+~ ( 454896 - 174960 \alpha ) \frac{\ln^2(3) \pi}{\sqrt{3}}
+ ( 2099520 \alpha - 5458752 ) \frac{\ln(3) \pi}{\sqrt{3}}
\right. \right. \nonumber \\
&& \left. \left. ~~~~~~~~~
+~ ( 187920 \alpha - 488592 ) \frac{\pi^3}{\sqrt{3}}
\right] C_A \right. \nonumber \\
&& ~~~~~+~ \left. \left[ 
2052864 \psi^\prime(\third) 
- 1368576 \pi^2 
- 32160888
\right] T_F \Nf 
\right] \frac{C_F a^2}{839808} \nonumber \\
&& +~ O(a^3) ~, 
\end{eqnarray} 
\begin{eqnarray}
C^{W_3}_{12}(a,\alpha) &=& \left[ ( 48 - 48 \alpha ) 
\psi^\prime(\third) + ( 32 \alpha - 32 ) \pi^2 + 36 \alpha - 927 \right] 
\frac{C_F a}{324} \nonumber \\
&& +~ \left[ \left[ 
( 276480 \alpha - 129024 \alpha^2 - 396288 ) (\psi^\prime(\third))^2 
\right. \right. \nonumber \\
&& \left. \left. ~~~~~
+~ ( 172032 \alpha^2 - 368640 \alpha + 528384 ) \psi^\prime(\third) \pi^2
\right. \right. \nonumber \\
&& \left. \left. ~~~~~
+~ ( 421632 \alpha^2 - 2847744 \alpha - 2817504 ) \psi^\prime(\third) 
- ( 155520 - 20736 \alpha ) \psi^{\prime\prime\prime}(\third) 
\right. \right. \nonumber \\
&& \left. \left. ~~~~~
-~ 73903104 s_2(\pisix) 
+ 147806208 s_2(\pitwo) 
+ 123171840 s_3(\pisix) 
- 98537472 s_3(\pitwo) 
\right. \right. \nonumber \\
&& \left. \left. ~~~~~
-~ ( 57344 \alpha^2 - 67584 \alpha - 238592 ) \pi^4 
\right. \right. \nonumber \\
&& \left. \left. ~~~~~
-~ ( 281088 \alpha^2 - 1898496 \alpha - 1878336 ) \pi^2 
- 383616 \alpha^2 + 440640 \alpha 
\right. \right. \nonumber \\
&& \left. \left. ~~~~~
-~ 17134821
- ( 373248 \alpha - 373248 ) \Sigma
+ 13996800 \zeta(3)
\right. \right. \nonumber \\
&& \left. \left. ~~~~~
-~ 513216 \frac{\ln^2(3) \pi}{\sqrt{3}}
+ 6158592 \frac{\ln(3) \pi}{\sqrt{3}}
+ 551232 \frac{\pi^3}{\sqrt{3}}
\right] C_F \right. \nonumber \\
&& ~~~~~+ \left. \left[ 
124416 (\psi^\prime(\third))^2 
- 165888 \psi^\prime(\third) \pi^2 
\right. \right. \nonumber \\
&& \left. \left. ~~~~~~~~~
-~ ( 207360 \alpha^2 + 147744 \alpha - 2208384 ) \psi^\prime(\third) 
\right. \right. \nonumber \\
&& \left. \left. ~~~~~~~~~
+~ ( 18144 + 15552 \alpha ) \psi^{\prime\prime\prime}(\third) 
+ ( 15116544 + 5038848 \alpha ) s_2(\pisix) 
\right. \right. \nonumber \\
&& \left. \left. ~~~~~~~~~
-~ ( 10077696 \alpha + 30233088 ) s_2(\pitwo) 
- ( 8398080 \alpha + 25194240 ) s_3(\pisix) 
\right. \right. \nonumber \\
&& \left. \left. ~~~~~~~~~
+~ ( 20155392 + 6718464 \alpha ) s_3(\pitwo) 
+ ( 6912 - 41472 \alpha ) \pi^4 
\right. \right. \nonumber \\
&& \left. \left. ~~~~~~~~~
+~ ( 138240 \alpha^2 + 98496 \alpha - 1472256 ) \pi^2 
+ 225504 \alpha^2 + 594864 \alpha 
\right. \right. \nonumber \\
&& \left. \left. ~~~~~~~~~
-~ 23258016
+ ( 93312 - 186624 \alpha ) \Sigma
+ ( 839808 - 839808 \alpha ) \zeta(3)
\right. \right. \nonumber \\
&& \left. \left. ~~~~~~~~~
+~ ( 34992 \alpha + 104976 ) \frac{\ln^2(3) \pi}{\sqrt{3}}
- ( 1259712 + 419904 \alpha ) \frac{\ln(3) \pi}{\sqrt{3}}
\right. \right. \nonumber \\
&& \left. \left. ~~~~~~~~~
-~ ( 112752+ 37584 \alpha ) \frac{\pi^3}{\sqrt{3}}
\right] C_A \right. \nonumber \\
&& ~~~~~+~ \left. \left[ 
373248 \pi^2 
- 559872 \psi^\prime(\third) 
+ 11611512
\right] T_F \Nf 
\right] \frac{C_F a^2}{839808} ~+~ O(a^3) ~, 
\end{eqnarray} 
\begin{eqnarray}
C^{W_3}_{13}(a,\alpha) &=& \left[ ( 48 \alpha - 24 ) \psi^\prime(\third) 
+ ( 16 - 32 \alpha ) \pi^2 - 198 \alpha - 9 \right] 
\frac{C_F a}{324} \nonumber \\
&& +~ \left[ \left[ 
( 129024 \alpha^2 - 193536 \alpha + 322560 ) (\psi^\prime(\third))^2 
\right. \right. \nonumber \\
&& \left. \left. ~~~~~
-~ ( 172032 \alpha^2 - 258048 \alpha + 430080 ) \psi^\prime(\third) \pi^2
\right. \right. \nonumber \\
&& \left. \left. ~~~~~
-~ ( 743040 \alpha^2 - 1156032 \alpha + 9328608 ) \psi^\prime(\third) 
- ( 134784 + 20736 \alpha ) \psi^{\prime\prime\prime}(\third) 
\right. \right. \nonumber \\
&& \left. \left. ~~~~~
-~ 80621568 s_2(\pisix) 
+ 161243136 s_2(\pitwo) 
+ 134369280 s_3(\pisix) 
- 107495424 s_3(\pitwo) 
\right. \right. \nonumber \\
&& \left. \left. ~~~~~
+~ ( 57344 \alpha^2 - 30720 \alpha + 502784 ) \pi^4 
\right. \right. \nonumber \\
&& \left. \left. ~~~~~
+~ ( 495360 \alpha^2 - 770688 \alpha + 6219072 ) \pi^2 
+ 974592 \alpha^2 + 629856 \alpha 
\right. \right. \nonumber \\
&& \left. \left. ~~~~~
-~ 2984931
+ ( 373248 \alpha - 186624 ) \Sigma
+ 11757312 \zeta(3)
\right. \right. \nonumber \\
&& \left. \left. ~~~~~
-~ 559872 \frac{\ln^2(3) \pi}{\sqrt{3}}
+ 6718464 \frac{\ln(3) \pi}{\sqrt{3}}
+ 601344 \frac{\pi^3}{\sqrt{3}}
\right] C_F \right. \nonumber \\
&& ~~~~~+ \left. \left[ 
165888 \psi^\prime(\third) \pi^2 
- 124416 (\psi^\prime(\third))^2 
\right. \right. \nonumber \\
&& \left. \left. ~~~~~~~~~
+~ ( 129600 \alpha^2 + 1314144 \alpha + 1428192 ) \psi^\prime(\third) 
\right. \right. \nonumber \\
&& \left. \left. ~~~~~~~~~
+~ ( 20736 + 15552 \alpha ) \psi^{\prime\prime\prime}(\third) 
+ ( 12597120 + 10077696 \alpha ) s_2(\pisix) 
\right. \right. \nonumber \\
&& \left. \left. ~~~~~~~~~
-~ ( 20155392 \alpha + 25194240 ) s_2(\pitwo) 
- ( 16796160 \alpha + 20995200 ) s_3(\pisix) 
\right. \right. \nonumber \\
&& \left. \left. ~~~~~~~~~
+~ ( 16796190 + 13436928 \alpha ) s_3(\pitwo) 
- ( 110592 + 41472 \alpha ) \pi^4 
\right. \right. \nonumber \\
&& \left. \left. ~~~~~~~~~
-~ ( 86400 \alpha^2 + 876096 \alpha + 952128 ) \pi^2 
- 517104 \alpha^2 - 1662120 \alpha 
\right. \right. \nonumber \\
&& \left. \left. ~~~~~~~~~
-~ 1840968
- ( 93312 - 186624 \alpha ) \Sigma
+ ( 629856 - 1679616 \alpha ) \zeta(3)
\right. \right. \nonumber \\
&& \left. \left. ~~~~~~~~~
+~ ( 69984 \alpha + 87480 ) \frac{\ln^2(3) \pi}{\sqrt{3}}
- ( 1049760 + 839808 \alpha ) \frac{\ln(3) \pi}{\sqrt{3}}
\right. \right. \nonumber \\
&& \left. \left. ~~~~~~~~~
-~ ( 93960+ 75168 \alpha ) \frac{\pi^3}{\sqrt{3}}
\right] C_A \right. \nonumber \\
&& ~~~~~+~ \left. \left[ 
311040 \psi^\prime(\third) 
- 207360 \pi^2 
+ 390744
\right] T_F \Nf 
\right] \frac{C_F a^2}{839808} ~+~ O(a^3) ~, 
\end{eqnarray} 
\begin{eqnarray}
C^{W_3}_{22}(a,\alpha) &=& 1 ~+~ \left[ ( 36 \alpha - 42 ) \psi^\prime(\third) 
+ ( 28 - 24 \alpha ) \pi^2 - 27 \alpha + 459 \right] \frac{C_F a}{81} 
\nonumber \\
&& +~ \left[ \left[ 
( 5184 \alpha^2 - 12096 \alpha - 4608 ) (\psi^\prime(\third))^2 
+ ( 16128 \alpha - 6912 \alpha^2 + 6144 ) \psi^\prime(\third) \pi^2
\right. \right. \nonumber \\
&& \left. \left. ~~~~~
+~ ( 328536 \alpha - 13608 \alpha^2 - 613656 ) \psi^\prime(\third) 
- ( 5022 + 1944 \alpha ) \psi^{\prime\prime\prime}(\third) 
\right. \right. \nonumber \\
&& \left. \left. ~~~~~
+~ ( 1259712 \alpha - 5248800 ) s_2(\pisix) 
+ ( 10497600 - 2519424 \alpha ) s_2(\pitwo) 
\right. \right. \nonumber \\
&& \left. \left. ~~~~~
+~ ( 8748000 - 2099520 \alpha ) s_3(\pisix) 
+ ( 1679616 \alpha - 6998400 ) s_3(\pitwo) 
\right. \right. \nonumber \\
&& \left. \left. ~~~~~
+~ ( 2304 \alpha^2 - 192 \alpha + 11344 ) \pi^4 
+ ( 9072 \alpha^2 - 219024 \alpha + 409104 ) \pi^2 
\right. \right. \nonumber \\
&& \left. \left. ~~~~~
+~ 7290 \alpha^2 - 90882 \alpha + 107568
+ ( 34992 \alpha - 40824 ) \Sigma
\right. \right. \nonumber \\
&& \left. \left. ~~~~~
+~ ( 104976 \alpha + 559872 ) \zeta(3)
+ ( 8748 \alpha - 36450 ) \frac{\ln^2(3) \pi}{\sqrt{3}}
\right. \right. \nonumber \\
&& \left. \left. ~~~~~
+~ ( 437400 - 104976 \alpha ) \frac{\ln(3) \pi}{\sqrt{3}}
+ ( 39150 - 9396 \alpha ) \frac{\pi^3}{\sqrt{3}}
\right] C_F \right. \nonumber \\
&& ~~~~~+ \left. \left[ 
5832 (\psi^\prime(\third))^2 
- 7776 \psi^\prime(\third) \pi^2 
+ ( 11664 \alpha^2 - 39366 \alpha + 99468 ) \psi^\prime(\third) 
\right. \right. \nonumber \\
&& \left. \left. ~~~~~~~~~
+~ ( 5103 - 972 \alpha ) \psi^{\prime\prime\prime}(\third) 
+ ( 2519424 - 629856 \alpha ) s_2(\pisix) 
\right. \right. \nonumber \\
&& \left. \left. ~~~~~~~~~
+~ ( 1259712 \alpha - 5038848 ) s_2(\pitwo) 
+ ( 1049760 \alpha - 4199040 ) s_3(\pisix) 
\right. \right. \nonumber \\
&& \left. \left. ~~~~~~~~~
+~ ( 3359232 - 839808 \alpha ) s_3(\pitwo) 
+ ( 2592 \alpha - 11016 ) \pi^4 
\right. \right. \nonumber \\
&& \left. \left. ~~~~~~~~~
-~ ( 7776 \alpha^2 - 26244 \alpha + 66312 ) \pi^2 
- 8748 \alpha^2 - 34992 \alpha + 1759644
\right. \right. \nonumber \\
&& \left. \left. ~~~~~~~~~
+~ ( 17496 \alpha - 34992 ) \Sigma
+ ( 26244 \alpha - 734832 ) \zeta(3)
\right. \right. \nonumber \\
&& \left. \left. ~~~~~~~~~
+~ ( 17496 - 4374 \alpha ) \frac{\ln^2(3) \pi}{\sqrt{3}}
+ ( 52488 \alpha - 209952 ) \frac{\ln(3) \pi}{\sqrt{3}}
\right. \right. \nonumber \\
&& \left. \left. ~~~~~~~~~
+~ ( 4698 \alpha - 18792 ) \frac{\pi^3}{\sqrt{3}}
\right] C_A \right. \nonumber \\
&& ~~~~~+~ \left. \left[ 
46656 \psi^\prime(\third) 
- 31104 \pi^2 
- 642168
\right] T_F \Nf 
\right] \frac{C_F a^2}{26244} ~+~ O(a^3) ~, 
\end{eqnarray} 
\begin{eqnarray}
C^{W_3}_{23}(a,\alpha) &=& \left[ 24 \psi^\prime(\third) 
- 16 \pi^2 - 54 \alpha - 297 \right] \frac{C_F a}{162} 
\nonumber \\
&& +~ \left[ \left[ 
( 13824 \alpha + 35136 ) (\psi^\prime(\third))^2 
- ( 18432 \alpha + 46848 ) \psi^\prime(\third) \pi^2
\right. \right. \nonumber \\
&& \left. \left. ~~~~~
-~ ( 342144 \alpha + 19440 \alpha^2 - 824904 ) \psi^\prime(\third) 
+ 18792 \psi^{\prime\prime\prime}(\third) 
\right. \right. \nonumber \\
&& \left. \left. ~~~~~
+~ ( 10917504 - 1679616 \alpha ) s_2(\pisix) 
+ ( 3359232 - 21835008 \alpha ) s_2(\pitwo) 
\right. \right. \nonumber \\
&& \left. \left. ~~~~~
+~ ( 2799360 - 18195840 \alpha ) s_3(\pisix) 
+ ( 14556672 - 2239488 \alpha ) s_3(\pitwo) 
\right. \right. \nonumber \\
&& \left. \left. ~~~~~
+~ ( 6144 \alpha - 34496 ) \pi^4 
+ ( 12960 \alpha^2 + 228096 \alpha - 549936 ) \pi^2 
\right. \right. \nonumber \\
&& \left. \left. ~~~~~
+~ 37908 \alpha^2 + 76788 \alpha - 188892
+ 46656 \Sigma
\right. \right. \nonumber \\
&& \left. \left. ~~~~~
-~ ( 139968 \alpha + 1609632 ) \zeta(3)
+ ( 75816 - 11664 \alpha ) \frac{\ln^2(3) \pi}{\sqrt{3}}
\right. \right. \nonumber \\
&& \left. \left. ~~~~~
+~ ( 139968 \alpha - 909792 ) \frac{\ln(3) \pi}{\sqrt{3}}
+ ( 12528 \alpha - 81432 ) \frac{\pi^3}{\sqrt{3}}
\right] C_F \right. \nonumber \\
&& ~~~~~+ \left. \left[ 
31104 \psi^\prime(\third) \pi^2 
- 23328 (\psi^\prime(\third))^2 
+ ( 107892 \alpha - 5832 \alpha^2 - 122472 ) \psi^\prime(\third) 
\right. \right. \nonumber \\
&& \left. \left. ~~~~~~~~~
+~ ( 972 \alpha - 9720 ) \psi^{\prime\prime\prime}(\third) 
+ ( 944784 \alpha - 4513968 ) s_2(\pisix) 
\right. \right. \nonumber \\
&& \left. \left. ~~~~~~~~~
+~ ( 9027936 - 1889568 \alpha ) s_2(\pitwo) 
+ ( 7523280 - 1574640 \alpha ) s_3(\pisix) 
\right. \right. \nonumber \\
&& \left. \left. ~~~~~~~~~
+~ ( 1259712 \alpha - 6018624 ) s_3(\pitwo) 
+ ( 15552 - 2592 \alpha ) \pi^4 
\right. \right. \nonumber \\
&& \left. \left. ~~~~~~~~~
+~ ( 3888 \alpha^2 - 71928 \alpha + 81648 ) \pi^2 
- 21870 \alpha^2 - 113724 \alpha - 2488482
\right. \right. \nonumber \\
&& \left. \left. ~~~~~~~~~
+~ 34992 \Sigma
+ ( 1277208 - 52488 \alpha ) \zeta(3)
\right. \right. \nonumber \\
&& \left. \left. ~~~~~~~~~
+~ ( 6561 \alpha - 31347 ) \frac{\ln^2(3) \pi}{\sqrt{3}}
+ ( 376164 - 78732 \alpha ) \frac{\ln(3) \pi}{\sqrt{3}}
\right. \right. \nonumber \\
&& \left. \left. ~~~~~~~~~
+~ ( 33669 - 7047 \alpha ) \frac{\pi^3}{\sqrt{3}}
\right] C_A \right. \nonumber \\
&& ~~~~~+~ \left. \left[ 
29376 \pi^2 
- 44064 \psi^\prime(\third) 
+ 946080
\right] T_F \Nf 
\right] \frac{C_F a^2}{104976} ~+~ O(a^3)
\end{eqnarray} 
and
\begin{equation}
C^{W_3}_{33}(a,\alpha) ~=~ 1 ~+~ O(a^3) ~. 
\end{equation} 
The final expression merely reflects the fact that $\partial \partial W_3$ is
related to the vector current which is a physical operator. The values for row
$2$ and $3$ correspond to the values of the conversion function matrix for
$W_2$ given in \cite{9}\footnote{The expression for $C^{W_3}_{23}(a,\alpha)$ 
corrects a typographical error in the presentation of the two loop term 
involving $C_F^2$ of the corresponding equation for $C^{W_2}_{12}(a,\alpha)$ 
recorded in \cite{9} but which was correct in the attached data file of that 
article.}. This is consistent with the Wilson operator renormalization being
part of a tower of operators. These expressions include the usual 
transcendental type of numbers such as powers of $\pi$ and the Riemann zeta 
function of odd argument as well as rationals. However, given the nature of the
subtraction point other classes of basic numbers arise. These include 
derivatives of the Euler $\psi$-function and the natural logarithm as well as 
the polylogarithm functions through the specific function
\begin{equation}
s_n(z) ~=~ \frac{1}{\sqrt{3}} \Im \left[ \mbox{Li}_n \left(
\frac{e^{iz}}{\sqrt{3}} \right) \right] ~.
\end{equation}
In addition a combination of various harmonic polylogarithms also appears which
is 
\begin{equation}
\Sigma ~=~ {\cal H}^{(2)}_{31} ~+~ {\cal H}^{(2)}_{43}
\end{equation}
using the notation of \cite{2}. These particular basic numbers derive from the
explicit expressions for the scalar master integrals which the Laporta 
algorithm produces as a result of the reduction of integrals. They have been
evaluated by various authors in \cite{43,44,45,46}. 

For the conversion to all the numerical forms we present here we have used the 
following numerical values for the various polygamma and polylogarithm 
functions which are 
\begin{eqnarray}
\zeta(3) &=& 1.20205690 ~~,~~ \Sigma ~=~ 6.34517334 ~~,~~
\psi^\prime\left( \frac{1}{3} \right) ~=~ 10.09559713 ~, \nonumber \\
\psi^{\prime\prime\prime}\left( \frac{1}{3} \right) &=& 488.1838167 ~~,~~
s_2\left( \frac{\pi}{2} \right) ~=~ 0.32225882 ~~,~~
s_2\left( \frac{\pi}{6} \right) ~=~ 0.22459602 ~, \nonumber \\
s_3\left( \frac{\pi}{2} \right) &=& 0.32948320 ~~,~~
s_3\left( \frac{\pi}{6} \right) ~=~ 0.19259341 ~.
\end{eqnarray}
Expressing the conversion matrix in numerical form for the colour group $SU(3)$
produces 
\begin{eqnarray}
C^{W_3}_{11}(a,\alpha) &=& 1 ~+~ [ 2.1853715 \alpha + 8.2451607 ] a \nonumber \\
&& +~ [ 9.9594216 \alpha^2 + 38.4241424 \alpha + 224.4265963 - 19.8007988 \Nf ]
a^2 ~+~ O(a^3) ~, \nonumber \\
C^{W_3}_{12}(a,\alpha) &=& -~ [ 0.5463429 \alpha + 3.1203238 ] a \nonumber \\
&& -~ [ 3.4486064 \alpha^2 + 15.2459694 \alpha + 91.8037478 - 7.6549878 \Nf ]
a^2 ~+~ O(a^3) ~, \nonumber \\
C^{W_3}_{13}(a,\alpha) &=& -~ [ 0.1203238 \alpha + 0.3842826 ] a \nonumber \\
&& -~ [ 0.3835601 \alpha^2 + 1.2332093 \alpha + 9.0543723 - 1.1782990 \Nf ]
a^2 ~+~ O(a^3) ~, \nonumber \\
C^{W_3}_{22}(a,\alpha) &=& 1 ~+~ [ 1.6390287 \alpha + 5.1248369 ] a \nonumber \\
&& +~ [ 6.5108151 \alpha^2 + 23.1781730 \alpha + 132.6228486 - 12.1458110 \Nf ]
a^2 ~+~ O(a^3) ~, \nonumber \\
C^{W_3}_{23}(a,\alpha) &=& -~ [ 0.4444444 \alpha + 1.7499534 ] a \nonumber \\
&& -~ [ 2.1301455 \alpha^2 + 7.2772405 \alpha + 49.1967063 - 5.0243682 \Nf ]
a^2 ~+~ O(a^3) ~, \nonumber \\
C^{W_3}_{33}(a,\alpha) &=& 1 ~+~ O(a^3) ~.
\end{eqnarray}
Clearly there is a large correction at two loops for $W_3$ itself compared with
$\partial W_3$ or $W_2$. As one of the motivations for developing the 
RI${}^\prime$/SMOM scheme was the hope that the convergence of the conversion
function would improve we recall the numerical values for the conversion
functions from RI${}^\prime$ to $\MSbar$. From \cite{47} we have the numerical
values in the Landau gauge  
\begin{eqnarray}
\tilde{C}^{W_3}_{11}(a,0) &=& 1 ~+~ 7.9259259 a ~+~ [ 215.0853593
- 18.9809671 \Nf ] a^2 ~+~ O(a^3) ~, \nonumber \\
\tilde{C}^{W_3}_{22}(a,0) &=& 1 ~+~ 4.5925926 a ~+~ [ 119.8268158
- 10.9794239 \Nf ] a^2 ~+~ O(a^3) ~, \nonumber \\
\tilde{C}^{W_3}_{33}(a,0) &=& 1 ~+~ O(a^3) ~.
\end{eqnarray}
Given the fact the top entry is the key one it appears that the 
RI${}^\prime$/SMOM scheme has a slightly larger two loop correction compared
to the RI${}^\prime$ conversion function. A similar observation was made for
$W_2$, \cite{10}. However, as noted in that article it is not entirely clear
whether one can truly compare these conversion functions. This is because the
nature of the RI${}^\prime$ scheme is such that it cannot access the 
off-diagonal elements of the mixing matrix. This stems from the momentum
configuration of the underlying Green's function. Indeed in this respect it is
not entirely clear whether one can regard RI${}^\prime$ as a full 
renormalization scheme for the Wilson operators as a result of the mixing with
total derivative operators. For operators where there is no mixing such as the
quark currents then a comparison of conversion functions would appear more
appropriate, \cite{1,2,3}.

\sect{Three loop anomalous dimensions.}

One aspect of the conversion functions is that they can be used to determine
the anomalous dimensions given knowledge of the anomalous dimensions in one
scheme. Although we are working with mixing matrices here it is straightforward
to extend the formalism to show that in our case
\begin{eqnarray}
\gamma^{W_3}_{ij,\mbox{\footnotesize{RI$^\prime$/SMOM}}}
\left(a_{\mbox{\footnotesize{RI$^\prime$}}},
\alpha_{\mbox{\footnotesize{RI$^\prime$}}}\right) &=&
C^{W_3}_{ik}\left(a_{\mbox{\footnotesize{$\MSbar$}}},
\alpha_{\mbox{\footnotesize{$\MSbar$}}}\right) 
\gamma^{W_3}_{kl,\mbox{\footnotesize{$\MSbar$}}}
\left(a_{\mbox{\footnotesize{$\MSbar$}}},
\alpha_{\mbox{\footnotesize{$\MSbar$}}}\right)
\left( C^{W_3}_{lj} \left(a_{\mbox{\footnotesize{$\MSbar$}}},
\alpha_{\mbox{\footnotesize{$\MSbar$}}}\right) \right)^{-1} \nonumber \\
&& -~ \left[ \mu \frac{d~}{d\mu} 
C^{W_3}_{ik}\left(a_{\mbox{\footnotesize{$\MSbar$}}},
\alpha_{\mbox{\footnotesize{$\MSbar$}}}\right) \right] 
\left( C^{W_3}_{kj} \left(a_{\mbox{\footnotesize{$\MSbar$}}},
\alpha_{\mbox{\footnotesize{$\MSbar$}}}\right) \right)^{-1} 
\end{eqnarray}
where we now have to explicitly label the scheme the variables correspond to by
the subscript. More explicitly for the three diagonal elements of the anomalous
dimension mixing matrix we have
\begin{eqnarray}
\gamma^{W_3}_{ii,\mbox{\footnotesize{RI$^\prime$/SMOM}}}
\left(a_{\mbox{\footnotesize{RI$^\prime$}}},
\alpha_{\mbox{\footnotesize{RI$^\prime$}}}\right) &=&
\gamma^{W_3}_{ii,\mbox{\footnotesize{$\MSbar$}}}
\left(a_{\mbox{\footnotesize{$\MSbar$}}}\right) ~-~
\beta\left(a_{\mbox{\footnotesize{$\MSbar$}}}\right)
\frac{\partial ~}{\partial a_{\mbox{\footnotesize{$\MSbar$}}}}
\ln C^{W_3}_{ii}\left(a_{\mbox{\footnotesize{$\MSbar$}}},
\alpha_{\mbox{\footnotesize{$\MSbar$}}}\right) \nonumber \\
&& -~ \alpha_{\mbox{\footnotesize{$\MSbar$}}}
\gamma^{\mbox{\footnotesize{$\MSbar$}}}_\alpha
\left(a_{\mbox{\footnotesize{$\MSbar$}}},
\alpha_{\mbox{\footnotesize{$\MSbar$}}}\right)
\frac{\partial ~}{\partial \alpha_{\mbox{\footnotesize{$\MSbar$}}}}
\ln C^{W_3}_{ii}\left(a_{\mbox{\footnotesize{$\MSbar$}}},
\alpha_{\mbox{\footnotesize{$\MSbar$}}}\right)
\end{eqnarray}
where again there is no sum over $i$ and $i$~$=$~$1$, $2$ or $3$. The 
off-diagonal elements are more involved since  
\begin{eqnarray}
\gamma^{W_3}_{12,\mbox{\footnotesize{RI$^\prime$/SMOM}}}
\left(a_{\mbox{\footnotesize{RI$^\prime$}}},
\alpha_{\mbox{\footnotesize{RI$^\prime$}}}\right) &=&
\left[ \gamma^{W_3}_{12,\mbox{\footnotesize{$\MSbar$}}}
\left(a_{\mbox{\footnotesize{$\MSbar$}}}\right)
C^{W_3}_{11}\left(a_{\mbox{\footnotesize{$\MSbar$}}},
\alpha_{\mbox{\footnotesize{$\MSbar$}}}\right) \right. \nonumber \\
&& \left. ~-
\beta\left(a_{\mbox{\footnotesize{$\MSbar$}}}\right)
\frac{\partial ~}{\partial a_{\mbox{\footnotesize{$\MSbar$}}}}
C^{W_3}_{12}\left(a_{\mbox{\footnotesize{$\MSbar$}}},
\alpha_{\mbox{\footnotesize{$\MSbar$}}}\right) \right. \nonumber \\
&& \left. ~- \alpha_{\mbox{\footnotesize{$\MSbar$}}}
\gamma^{\mbox{\footnotesize{$\MSbar$}}}_\alpha
\left(a_{\mbox{\footnotesize{$\MSbar$}}},
\alpha_{\mbox{\footnotesize{$\MSbar$}}}\right)
\frac{\partial ~}{\partial \alpha_{\mbox{\footnotesize{$\MSbar$}}}}
C^{W_3}_{12}\left(a_{\mbox{\footnotesize{$\MSbar$}}},
\alpha_{\mbox{\footnotesize{$\MSbar$}}}\right) \right. \nonumber \\
&& \left. ~- \gamma^{W_3}_{11,\mbox{\footnotesize{$\MSbar$}}}
\left(a_{\mbox{\footnotesize{$\MSbar$}}}\right)
C^{W_3}_{12}\left(a_{\mbox{\footnotesize{$\MSbar$}}},
\alpha_{\mbox{\footnotesize{$\MSbar$}}}\right) \right. \nonumber \\
&& \left. ~+ \gamma^{W_3}_{22,\mbox{\footnotesize{$\MSbar$}}}
\left(a_{\mbox{\footnotesize{$\MSbar$}}}\right)
C^{W_3}_{12}\left(a_{\mbox{\footnotesize{$\MSbar$}}},
\alpha_{\mbox{\footnotesize{$\MSbar$}}}\right) \right. \nonumber \\
&& \left. ~+ C^{W_3}_{12}\left(a_{\mbox{\footnotesize{$\MSbar$}}},
\alpha_{\mbox{\footnotesize{$\MSbar$}}}\right)
\beta\left(a_{\mbox{\footnotesize{$\MSbar$}}}\right)
\frac{\partial ~}{\partial a_{\mbox{\footnotesize{$\MSbar$}}}}
\ln C^{W_3}_{11}\left(a_{\mbox{\footnotesize{$\MSbar$}}},
\alpha_{\mbox{\footnotesize{$\MSbar$}}}\right) \right. \nonumber \\
&& \left. ~+
C^{W_3}_{12}\!\left(a_{\mbox{\footnotesize{$\MSbar$}}},
\alpha_{\mbox{\footnotesize{$\MSbar$}}}\right)
\alpha_{\mbox{\footnotesize{$\MSbar$}}}
\gamma^{\mbox{\footnotesize{$\MSbar$}}}_\alpha
\left(a_{\mbox{\footnotesize{$\MSbar$}}},
\alpha_{\mbox{\footnotesize{$\MSbar$}}}\right) \right. \nonumber \\
&& \left. ~~~~\times
\frac{\partial ~}{\partial \alpha_{\mbox{\footnotesize{$\MSbar$}}}}
\ln C^{W_3}_{11}\left(a_{\mbox{\footnotesize{$\MSbar$}}},
\alpha_{\mbox{\footnotesize{$\MSbar$}}}\right) \right]
\left[ C^{W_3}_{22}\left(a_{\mbox{\footnotesize{$\MSbar$}}},
\alpha_{\mbox{\footnotesize{$\MSbar$}}}\right)
\right]^{-1} \!\!,
\end{eqnarray}
\begin{eqnarray}
\gamma^{W_3}_{13,\mbox{\footnotesize{RI$^\prime$/SMOM}}}
\left(a_{\mbox{\footnotesize{RI$^\prime$}}},
\alpha_{\mbox{\footnotesize{RI$^\prime$}}}\right) &=&
\left[ \gamma^{W_3}_{13,\mbox{\footnotesize{$\MSbar$}}}
\left(a_{\mbox{\footnotesize{$\MSbar$}}}\right)
C^{W_3}_{11}\left(a_{\mbox{\footnotesize{$\MSbar$}}},
\alpha_{\mbox{\footnotesize{$\MSbar$}}}\right) \right. \nonumber \\
&& \left. ~+ \gamma^{W_3}_{23,\mbox{\footnotesize{$\MSbar$}}}
\left(a_{\mbox{\footnotesize{$\MSbar$}}}\right)
C^{W_3}_{12}\left(a_{\mbox{\footnotesize{$\MSbar$}}},
\alpha_{\mbox{\footnotesize{$\MSbar$}}}\right) \right. \nonumber \\
&& \left. ~- \gamma^{W_3}_{11,\mbox{\footnotesize{$\MSbar$}}}
\left(a_{\mbox{\footnotesize{$\MSbar$}}}\right)
C^{W_3}_{13}\left(a_{\mbox{\footnotesize{$\MSbar$}}},
\alpha_{\mbox{\footnotesize{$\MSbar$}}}\right) \right. \nonumber \\
&& \left. ~+ \gamma^{W_3}_{33,\mbox{\footnotesize{$\MSbar$}}}
\left(a_{\mbox{\footnotesize{$\MSbar$}}}\right)
C^{W_3}_{13}\left(a_{\mbox{\footnotesize{$\MSbar$}}},
\alpha_{\mbox{\footnotesize{$\MSbar$}}}\right) \right. \nonumber \\
&& \left. ~-
\beta\left(a_{\mbox{\footnotesize{$\MSbar$}}}\right)
\frac{\partial ~}{\partial a_{\mbox{\footnotesize{$\MSbar$}}}}
C^{W_3}_{13}\left(a_{\mbox{\footnotesize{$\MSbar$}}},
\alpha_{\mbox{\footnotesize{$\MSbar$}}}\right) \right. \nonumber \\
&& \left. ~- \alpha_{\mbox{\footnotesize{$\MSbar$}}}
\gamma^{\mbox{\footnotesize{$\MSbar$}}}_\alpha
\left(a_{\mbox{\footnotesize{$\MSbar$}}},
\alpha_{\mbox{\footnotesize{$\MSbar$}}}\right)
\frac{\partial ~}{\partial \alpha_{\mbox{\footnotesize{$\MSbar$}}}}
C^{W_3}_{13}\left(a_{\mbox{\footnotesize{$\MSbar$}}},
\alpha_{\mbox{\footnotesize{$\MSbar$}}}\right) \right. \nonumber \\
&& \left. ~+ C^{W_3}_{13}\left(a_{\mbox{\footnotesize{$\MSbar$}}},
\alpha_{\mbox{\footnotesize{$\MSbar$}}}\right)
\beta\left(a_{\mbox{\footnotesize{$\MSbar$}}}\right)
\frac{\partial ~}{\partial a_{\mbox{\footnotesize{$\MSbar$}}}}
\ln C^{W_3}_{11}\left(a_{\mbox{\footnotesize{$\MSbar$}}},
\alpha_{\mbox{\footnotesize{$\MSbar$}}}\right) \right. \nonumber \\
&& \left. ~+
C^{W_3}_{13}\!\left(a_{\mbox{\footnotesize{$\MSbar$}}},
\alpha_{\mbox{\footnotesize{$\MSbar$}}}\right)
\alpha_{\mbox{\footnotesize{$\MSbar$}}}
\gamma^{\mbox{\footnotesize{$\MSbar$}}}_\alpha
\left(a_{\mbox{\footnotesize{$\MSbar$}}},
\alpha_{\mbox{\footnotesize{$\MSbar$}}}\right) \right. \nonumber \\
&& \left. ~~~~\times
\frac{\partial ~}{\partial \alpha_{\mbox{\footnotesize{$\MSbar$}}}}
\ln C^{W_3}_{11}\left(a_{\mbox{\footnotesize{$\MSbar$}}},
\alpha_{\mbox{\footnotesize{$\MSbar$}}}\right) \right]
\left[ C^{W_3}_{33}\left(a_{\mbox{\footnotesize{$\MSbar$}}},
\alpha_{\mbox{\footnotesize{$\MSbar$}}}\right)
\right]^{-1} \nonumber \\
&& +~ \left[ \gamma^{W_3}_{11,\mbox{\footnotesize{$\MSbar$}}}
\left(a_{\mbox{\footnotesize{$\MSbar$}}}\right)
C^{W_3}_{12}\left(a_{\mbox{\footnotesize{$\MSbar$}}},
\alpha_{\mbox{\footnotesize{$\MSbar$}}}\right) \right. \nonumber \\
&& \left. ~~~~~- \gamma^{W_3}_{12,\mbox{\footnotesize{$\MSbar$}}}
\left(a_{\mbox{\footnotesize{$\MSbar$}}}\right)
C^{W_3}_{11}\left(a_{\mbox{\footnotesize{$\MSbar$}}},
\alpha_{\mbox{\footnotesize{$\MSbar$}}}\right) \right. \nonumber \\
&& \left. ~~~~~- \gamma^{W_3}_{22,\mbox{\footnotesize{$\MSbar$}}}
\left(a_{\mbox{\footnotesize{$\MSbar$}}}\right)
C^{W_3}_{12}\left(a_{\mbox{\footnotesize{$\MSbar$}}},
\alpha_{\mbox{\footnotesize{$\MSbar$}}}\right) \right. \nonumber \\
&& \left. ~~~~~+
\beta\left(a_{\mbox{\footnotesize{$\MSbar$}}}\right)
\frac{\partial ~}{\partial a_{\mbox{\footnotesize{$\MSbar$}}}}
C^{W_3}_{12}\left(a_{\mbox{\footnotesize{$\MSbar$}}},
\alpha_{\mbox{\footnotesize{$\MSbar$}}}\right) \right. \nonumber \\
&& \left. ~~~~~+ \alpha_{\mbox{\footnotesize{$\MSbar$}}}
\gamma^{\mbox{\footnotesize{$\MSbar$}}}_\alpha
\left(a_{\mbox{\footnotesize{$\MSbar$}}},
\alpha_{\mbox{\footnotesize{$\MSbar$}}}\right)
\frac{\partial ~}{\partial \alpha_{\mbox{\footnotesize{$\MSbar$}}}}
C^{W_3}_{12}\left(a_{\mbox{\footnotesize{$\MSbar$}}},
\alpha_{\mbox{\footnotesize{$\MSbar$}}}\right) \right. \nonumber \\
&& \left. ~~~~~- C^{W_3}_{12}\left(a_{\mbox{\footnotesize{$\MSbar$}}},
\alpha_{\mbox{\footnotesize{$\MSbar$}}}\right)
\beta\left(a_{\mbox{\footnotesize{$\MSbar$}}}\right)
\frac{\partial ~}{\partial a_{\mbox{\footnotesize{$\MSbar$}}}}
\ln C^{W_3}_{11}\left(a_{\mbox{\footnotesize{$\MSbar$}}},
\alpha_{\mbox{\footnotesize{$\MSbar$}}}\right) \right. \nonumber \\
&& \left. ~~~~~-
C^{W_3}_{12}\!\left(a_{\mbox{\footnotesize{$\MSbar$}}},
\alpha_{\mbox{\footnotesize{$\MSbar$}}}\right)
\alpha_{\mbox{\footnotesize{$\MSbar$}}}
\gamma^{\mbox{\footnotesize{$\MSbar$}}}_\alpha
\left(a_{\mbox{\footnotesize{$\MSbar$}}},
\alpha_{\mbox{\footnotesize{$\MSbar$}}}\right) \right. \nonumber \\
&& \left. ~~~~~~~~\times
\frac{\partial ~}{\partial \alpha_{\mbox{\footnotesize{$\MSbar$}}}}
\ln C^{W_3}_{11}\left(a_{\mbox{\footnotesize{$\MSbar$}}},
\alpha_{\mbox{\footnotesize{$\MSbar$}}}\right) \right]
C^{W_3}_{23}\left(a_{\mbox{\footnotesize{$\MSbar$}}},
\alpha_{\mbox{\footnotesize{$\MSbar$}}}\right) \nonumber \\
&& ~~~~ \times \left[ C^{W_3}_{22}\left(a_{\mbox{\footnotesize{$\MSbar$}}},
\alpha_{\mbox{\footnotesize{$\MSbar$}}}\right)
C^{W_3}_{33}\left(a_{\mbox{\footnotesize{$\MSbar$}}},
\alpha_{\mbox{\footnotesize{$\MSbar$}}}\right)
\right]^{-1} ~,  
\end{eqnarray}
and 
\begin{eqnarray}
\gamma^{W_3}_{23,\mbox{\footnotesize{RI$^\prime$/SMOM}}}
\left(a_{\mbox{\footnotesize{RI$^\prime$}}},
\alpha_{\mbox{\footnotesize{RI$^\prime$}}}\right) &=&
\left[ \gamma^{W_3}_{23,\mbox{\footnotesize{$\MSbar$}}}
\left(a_{\mbox{\footnotesize{$\MSbar$}}}\right)
C^{W_3}_{22}\left(a_{\mbox{\footnotesize{$\MSbar$}}},
\alpha_{\mbox{\footnotesize{$\MSbar$}}}\right) \right. \nonumber \\
&& \left. ~-
\beta\left(a_{\mbox{\footnotesize{$\MSbar$}}}\right)
\frac{\partial ~}{\partial a_{\mbox{\footnotesize{$\MSbar$}}}}
C^{W_3}_{23}\left(a_{\mbox{\footnotesize{$\MSbar$}}},
\alpha_{\mbox{\footnotesize{$\MSbar$}}}\right) \right. \nonumber \\
&& \left. ~- \alpha_{\mbox{\footnotesize{$\MSbar$}}}
\gamma^{\mbox{\footnotesize{$\MSbar$}}}_\alpha
\left(a_{\mbox{\footnotesize{$\MSbar$}}},
\alpha_{\mbox{\footnotesize{$\MSbar$}}}\right)
\frac{\partial ~}{\partial \alpha_{\mbox{\footnotesize{$\MSbar$}}}}
C^{W_3}_{23}\left(a_{\mbox{\footnotesize{$\MSbar$}}},
\alpha_{\mbox{\footnotesize{$\MSbar$}}}\right) \right. \nonumber \\
&& \left. ~- \gamma^{W_3}_{22,\mbox{\footnotesize{$\MSbar$}}}
\left(a_{\mbox{\footnotesize{$\MSbar$}}}\right)
C^{W_3}_{23}\left(a_{\mbox{\footnotesize{$\MSbar$}}},
\alpha_{\mbox{\footnotesize{$\MSbar$}}}\right) \right. \nonumber \\
&& \left. ~+ \gamma^{W_3}_{33,\mbox{\footnotesize{$\MSbar$}}}
\left(a_{\mbox{\footnotesize{$\MSbar$}}}\right)
C^{W_3}_{23}\left(a_{\mbox{\footnotesize{$\MSbar$}}},
\alpha_{\mbox{\footnotesize{$\MSbar$}}}\right) \right. \nonumber \\
&& \left. ~+ C^{W_3}_{23}\left(a_{\mbox{\footnotesize{$\MSbar$}}},
\alpha_{\mbox{\footnotesize{$\MSbar$}}}\right)
\beta\left(a_{\mbox{\footnotesize{$\MSbar$}}}\right)
\frac{\partial ~}{\partial a_{\mbox{\footnotesize{$\MSbar$}}}}
\ln C^{W_3}_{22}\left(a_{\mbox{\footnotesize{$\MSbar$}}},
\alpha_{\mbox{\footnotesize{$\MSbar$}}}\right) \right. \nonumber \\
&& \left. ~+
C^{W_3}_{23}\!\left(a_{\mbox{\footnotesize{$\MSbar$}}},
\alpha_{\mbox{\footnotesize{$\MSbar$}}}\right)
\alpha_{\mbox{\footnotesize{$\MSbar$}}}
\gamma^{\mbox{\footnotesize{$\MSbar$}}}_\alpha
\left(a_{\mbox{\footnotesize{$\MSbar$}}},
\alpha_{\mbox{\footnotesize{$\MSbar$}}}\right) \right. \nonumber \\
&& \left. ~~~~\times
\frac{\partial ~}{\partial \alpha_{\mbox{\footnotesize{$\MSbar$}}}}
\ln C^{W_3}_{22}\left(a_{\mbox{\footnotesize{$\MSbar$}}},
\alpha_{\mbox{\footnotesize{$\MSbar$}}}\right) \right]
\left[ C^{W_3}_{33}\left(a_{\mbox{\footnotesize{$\MSbar$}}},
\alpha_{\mbox{\footnotesize{$\MSbar$}}}\right) \right]^{-1} \!\! .
\end{eqnarray}
Using these we have first verified that the two loop arbitrary gauge anomalous
dimensions, (\ref{andim2}), are reproduced. However, equipped with the explicit
values of the two loop conversion functions as well as various three loop 
$\MSbar$ anomalous dimensions we have the Landau gauge expressions  
\begin{eqnarray}
\left. \frac{}{} \gamma^{W_3}_{11}(a,0) 
\right|_{\mbox{\footnotesize{RI$^\prime$/SMOM}}} &=& \frac{25}{6} C_F a 
\nonumber \\
&& +~ \left[ \left[ 704 \pi^2 - 1056 \psi^\prime(\third) + 22068 
\right] C_A ~-~ 2035 C_F \right. \nonumber \\
&& \left. ~~~~+~ \left[ 384 \psi^\prime(\third) - 256 \pi^2 - 8232 \right]
T_F \Nf \right] \frac{C_F a^2}{432} ~+~ O(a^3) \nonumber \\
&& +~ \left[ 
\left[
25344 (\psi^\prime(\third))^2 
- 33792 \psi^\prime(\third) \pi^2
+ 44544 \psi^\prime(\third)
\right. \right. \nonumber \\
&& \left. \left. ~~~~~
+~ 59136 \psi^{\prime\prime\prime}(\third) 
+ 26687232 s_2(\pisix) 
- 53374464 s_2(\pitwo) 
\right. \right. \nonumber \\
&& \left. \left. ~~~~~
-~ 44478720 s_3(\pisix) 
+ 35582976 s_3(\pitwo) 
- 146432 \pi^4 
- 29696 \pi^2
\right. \right. \nonumber \\
&& \left. \left. ~~~~~
-~ 494208 \Sigma
- 9066816 \zeta(3)
+ 42261846
+ 185328 \frac{\ln^2(3) \pi}{\sqrt{3}}
\right. \right. \nonumber \\
&& \left. \left. ~~~~~
-~ 2223936 \frac{\ln(3) \pi}{\sqrt{3}}
- 199056 \frac{\pi^3}{\sqrt{3}}
\right] C_A^2 \right. \nonumber \\
&& \left. ~~~~+ \left[
25344 (\psi^\prime(\third))^2 
- 33792 \psi^\prime(\third) \pi^2
- 4907232 \psi^\prime(\third) 
\right. \right. \nonumber \\
&& \left. \left. ~~~~~~~~
-~ 2112 \psi^{\prime\prime\prime}(\third) 
- 38320128 s_2(\pisix) 
+ 76640256 s_2(\pitwo) 
\right. \right. \nonumber \\
&& \left. \left. ~~~~~~~~
+~ 63866880 s_3(\pisix) 
- 51093504 s_3(\pitwo) 
+ 16896 \pi^4 
\right. \right. \nonumber \\
&& \left. \left. ~~~~~~~~
+~ 3271488 \pi^2
- 684288 \Sigma
- 969408 \zeta(3)
- 4991286
\right. \right. \nonumber \\
&& \left. \left. ~~~~~~~~
-~ 266112 \frac{\ln^2(3) \pi}{\sqrt{3}}
+ 3193344 \frac{\ln(3) \pi}{\sqrt{3}}
+ 285824 \frac{\pi^3}{\sqrt{3}}
\right] C_A C_F \right. \nonumber \\
&& \left. ~~~~+ \left[
12288 \psi^\prime(\third) \pi^2
- 9216 (\psi^\prime(\third))^2 
+ 899328 \psi^\prime(\third) 
\right. \right. \nonumber \\
&& \left. \left. ~~~~~~~~
-~ 21504 \psi^{\prime\prime\prime}(\third) 
- 9704448 s_2(\pisix) 
+ 19408896 s_2(\pitwo) 
\right. \right. \nonumber \\
&& \left. \left. ~~~~~~~~
+~ 16174080 s_3(\pisix) 
- 12939264 s_3(\pitwo) 
+ 53248 \pi^4 
\right. \right. \nonumber \\
&& \left. \left. ~~~~~~~~
-~ 599552 \pi^2
+ 179712 \Sigma
+ 497664 \zeta(3)
- 29036688
\right. \right. \nonumber \\
&& \left. \left. ~~~~~~~~
-~ 67392 \frac{\ln^2(3) \pi}{\sqrt{3}}
+ 808704 \frac{\ln(3) \pi}{\sqrt{3}}
+ 72384 \frac{\pi^3}{\sqrt{3}}
\right] C_A T_F \Nf \right. \nonumber \\
&& \left. ~~~~+ \left[
12288 \psi^\prime(\third) \pi^2
- 9216 (\psi^\prime(\third))^2 
+ 1908864 \psi^\prime(\third) 
\right. \right. \nonumber \\
&& \left. \left. ~~~~~~~~
+~ 768 \psi^{\prime\prime\prime}(\third) 
+ 13934592 s_2(\pisix) 
- 27869184 s_2(\pitwo) 
\right. \right. \nonumber \\
&& \left. \left. ~~~~~~~~
-~ 23224320 s_3(\pisix) 
+ 18579456 s_3(\pitwo) 
- 6144 \pi^4 
\right. \right. \nonumber \\
&& \left. \left. ~~~~~~~~
-~ 1272576 \pi^2
+ 248832 \Sigma
+ 2529792 \zeta(3)
- 2559504
\right. \right. \nonumber \\
&& \left. \left. ~~~~~~~~
+\, 96768 \frac{\ln^2(3) \pi}{\sqrt{3}}
- 1161216 \frac{\ln(3) \pi}{\sqrt{3}}
- 103936 \frac{\pi^3}{\sqrt{3}}
\right] C_F T_F \Nf \right. \nonumber \\
&& \left. ~~~~+ \left[
202752 \pi^2 
- 304128 \psi^\prime(\third) 
+ 4517952
\right] T_F^2 \Nf^2 \right. \nonumber \\
&& \left. ~~~~+ \left[
1710720 \zeta(3) 
- 733515
\right] C_F^2 \right] \frac{C_F a^3}{46656} ~+~ O(a^4) ~, 
\end{eqnarray} 
\begin{eqnarray}
\left. \frac{}{} \gamma^{W_3}_{22}(a,0) 
\right|_{\mbox{\footnotesize{RI$^\prime$/SMOM}}} &=& 
\frac{8}{3} C_F a 
\nonumber \\
&& +~ \left[ \left[ 308 \pi^2 - 462 \psi^\prime(\third) + 8433 \right] C_A ~-~
1008 C_F \right. \nonumber \\
&& \left. ~~~~+~ \left[ 168 \psi^\prime(\third) - 112 \pi^2 - 2988 \right]
T_F \Nf \right] \frac{C_F a^2}{243} ~+~ O(a^3) \nonumber \\
&& +~ \left[ 
\left[
64152 (\psi^\prime(\third))^2 
- 85536 \psi^\prime(\third) \pi^2
+ 862812 \psi^\prime(\third)
\right. \right. \nonumber \\
&& \left. \left. ~~~~~
+~ 56133 \psi^{\prime\prime\prime}(\third) 
+ 27713664 s_2(\pisix) 
- 55427328 s_2(\pitwo) 
\right. \right. \nonumber \\
&& \left. \left. ~~~~~
-~ 46189440 s_3(\pisix) 
+ 36951552 s_3(\pitwo) 
- 121176 \pi^4 
- 575208 \pi^2
\right. \right. \nonumber \\
&& \left. \left. ~~~~~
-~ 384912 \Sigma
- 7243344 \zeta(3)
+ 25273296
+ 192456 \frac{\ln^2(3) \pi}{\sqrt{3}}
\right. \right. \nonumber \\
&& \left. \left. ~~~~~
-~ 2309472 \frac{\ln(3) \pi}{\sqrt{3}}
- 206712 \frac{\pi^3}{\sqrt{3}}
\right] C_A^2 \right. \nonumber \\
&& \left. ~~~~+ \left[
119328 \psi^\prime(\third) \pi^2
- 89496 (\psi^\prime(\third))^2 
- 5901984 \psi^\prime(\third) 
\right. \right. \nonumber \\
&& \left. \left. ~~~~~~~~
-~ 55242 \psi^{\prime\prime\prime}(\third) 
- 57736800 s_2(\pisix) 
+ 115473600 s_2(\pitwo) 
\right. \right. \nonumber \\
&& \left. \left. ~~~~~~~~
+~ 96228000 s_3(\pisix) 
- 76982400 s_3(\pitwo) 
+ 107536 \pi^4 
\right. \right. \nonumber \\
&& \left. \left. ~~~~~~~~
+~ 3934656 \pi^2
- 449064 \Sigma
+ 3639168 \zeta(3)
- 4833270
\right. \right. \nonumber \\
&& \left. \left. ~~~~~~~~
-~ 400950 \frac{\ln^2(3) \pi}{\sqrt{3}}
+ 4811400 \frac{\ln(3) \pi}{\sqrt{3}}
+ 430650 \frac{\pi^3}{\sqrt{3}}
\right] C_A C_F \right. \nonumber \\
&& \left. ~~~~+ \left[
31104 (\psi^\prime(\third))^2 
- 23328 \psi^\prime(\third) \pi^2
+ 251424 \psi^\prime(\third) 
\right. \right. \nonumber \\
&& \left. \left. ~~~~~~~~
-~ 20412 \psi^{\prime\prime\prime}(\third) 
- 10077696 s_2(\pisix) 
+ 20155392 s_2(\pitwo) 
\right. \right. \nonumber \\
&& \left. \left. ~~~~~~~~
+~ 16796160 s_3(\pisix) 
- 13436928 s_3(\pitwo) 
+ 44064 \pi^4 
\right. \right. \nonumber \\
&& \left. \left. ~~~~~~~~
-~ 167616 \pi^2
+ 139968 \Sigma
+ 1259712 \zeta(3)
- 16603056
\right. \right. \nonumber \\
&& \left. \left. ~~~~~~~~
-~ 69984 \frac{\ln^2(3) \pi}{\sqrt{3}}
+ 839808 \frac{\ln(3) \pi}{\sqrt{3}}
+ 75168 \frac{\pi^3}{\sqrt{3}}
\right] C_A T_F \Nf \right. \nonumber \\
&& \left. ~~~~+ \left[
32544 (\psi^\prime(\third))^2 
- 43392 \psi^\prime(\third) \pi^2
+ 2227824 \psi^\prime(\third) 
\right. \right. \nonumber \\
&& \left. \left. ~~~~~~~~
+~ 20088 \psi^{\prime\prime\prime}(\third) 
+ 20995200 s_2(\pisix) 
- 41990400 s_2(\pitwo) 
\right. \right. \nonumber \\
&& \left. \left. ~~~~~~~~
-~ 34992000 s_3(\pisix) 
+ 27993600 s_3(\pitwo) 
- 39104 \pi^4 
\right. \right. \nonumber \\
&& \left. \left. ~~~~~~~~
-~ 1485216 \pi^2
+ 163296 \Sigma
- 559872 \zeta(3)
- 742608
\right. \right. \nonumber \\
&& \left. \left. ~~~~~~~~
+\, 145800 \frac{\ln^2(3) \pi}{\sqrt{3}}
- 1749600 \frac{\ln(3) \pi}{\sqrt{3}}
- 156600 \frac{\pi^3}{\sqrt{3}}
\right] C_F T_F \Nf \right. \nonumber \\
&& \left. ~~~~+ \left[
124416 \pi^2 
- 186624 \psi^\prime(\third) 
+ 2423520 
\right] T_F^2 \Nf^2 \right. \nonumber \\
&& \left. ~~~~+ \left[
1679616 \zeta(3) 
- 90720 
\right] C_F^2 \right] \frac{C_F a^3}{39366} ~+~ O(a^4) ~, 
\end{eqnarray} 
\begin{eqnarray}
\left. \frac{}{} \gamma^{W_3}_{23}(a,0) 
\right|_{\mbox{\footnotesize{RI$^\prime$/SMOM}}} &=& 
-~ \frac{4}{3} C_F a 
\nonumber \\
&& +~ \left[ \left[ 264 \psi^\prime(\third) - 176 \pi^2 - 6651 \right] C_A ~+~
\left[ 144 \psi^\prime(\third) - 96 \pi^2 - 288 \right] C_F \right. 
\nonumber \\
&& \left. ~~~~+~ \left[ 64 \pi^2 - 96 \psi^\prime(\third) + 2340 \right]
T_F \Nf \right] \frac{C_F a^2}{486} ~+~ O(a^3) \nonumber \\
&& +~ \left[ 
\left[
342144 \psi^\prime(\third) \pi^2
- 256608 (\psi^\prime(\third))^2 
- 1082808 \psi^\prime(\third) 
\right. \right. \nonumber \\
&& \left. \left. ~~~~~
-~ 106920 \psi^{\prime\prime\prime}(\third) 
- 49653648 s_2(\pisix) 
+ 99307296 s_2(\pitwo) 
\right. \right. \nonumber \\
&& \left. \left. ~~~~~
+~ 82756080 s_3(\pisix) 
- 66204864 s_3(\pitwo) 
+ 171072 \pi^4 
+ 721872 \pi^2
\right. \right. \nonumber \\
&& \left. \left. ~~~~~
+~ 384912 \Sigma
+ 12369672 \zeta(3)
- 37423134
- 344817 \frac{\ln^2(3) \pi}{\sqrt{3}}
\right. \right. \nonumber \\
&& \left. \left. ~~~~~
+~ 4137804 \frac{\ln(3) \pi}{\sqrt{3}}
+ 370359 \frac{\pi^3}{\sqrt{3}}
\right] C_A^2 \right. \nonumber \\
&& \left. ~~~~+ \left[
477504 (\psi^\prime(\third))^2 
- 636672 \psi^\prime(\third) \pi^2
+ 7978176 \psi^\prime(\third) 
\right. \right. \nonumber \\
&& \left. \left. ~~~~~~~~
+~ 204768 \psi^{\prime\prime\prime}(\third) 
+ 117993024 s_2(\pisix) 
- 235986048 s_2(\pitwo) 
\right. \right. \nonumber \\
&& \left. \left. ~~~~~~~~
-~ 196655040 s_3(\pisix) 
+ 157324032 s_3(\pitwo) 
- 333824 \pi^4 
\right. \right. \nonumber \\
&& \left. \left. ~~~~~~~~
-~ 5318784 \pi^2
+ 653184 \Sigma
- 11897280 \zeta(3)
+ 367416
\right. \right. \nonumber \\
&& \left. \left. ~~~~~~~~
+~ 819396 \frac{\ln^2(3) \pi}{\sqrt{3}}
- 9832752 \frac{\ln(3) \pi}{\sqrt{3}}
- 880092 \frac{\pi^3}{\sqrt{3}}
\right] C_A C_F \right. \nonumber \\
&& \left. ~~~~+ \left[
93312 (\psi^\prime(\third))^2 
- 124416 \psi^\prime(\third) \pi^2
- 150336 \psi^\prime(\third) 
\right. \right. \nonumber \\
&& \left. \left. ~~~~~~~~
+~ 38880 \psi^{\prime\prime\prime}(\third) 
+ 18055872 s_2(\pisix) 
- 36111744 s_2(\pitwo) 
\right. \right. \nonumber \\
&& \left. \left. ~~~~~~~~
-~ 30093120 s_3(\pisix) 
+ 24074496 s_3(\pitwo) 
- 62208 \pi^4 
\right. \right. \nonumber \\
&& \left. \left. ~~~~~~~~
+~ 100224 \pi^2
- 139968 \Sigma
- 1749600 \zeta(3)
+ 24312312
\right. \right. \nonumber \\
&& \left. \left. ~~~~~~~~
+\, 125388 \frac{\ln^2(3) \pi}{\sqrt{3}}
- 1504656 \frac{\ln(3) \pi}{\sqrt{3}}
- 134676 \frac{\pi^3}{\sqrt{3}}
\right] C_A T_F \Nf \right. \nonumber \\
&& \left. ~~~~+ \left[
208896 \psi^\prime(\third) \pi^2
- 156672 (\psi^\prime(\third))^2 
- 3297024 \psi^\prime(\third) 
\right. \right. \nonumber \\
&& \left. \left. ~~~~~~~~
-~ 75168 \psi^{\prime\prime\prime}(\third) 
- 43670016 s_2(\pisix) 
+ 87340032 s_2(\pitwo) 
\right. \right. \nonumber \\
&& \left. \left. ~~~~~~~~
+~ 72783360 s_3(\pisix) 
- 58226688 s_3(\pitwo) 
+ 130816 \pi^4 
\right. \right. \nonumber \\
&& \left. \left. ~~~~~~~~
+~ 2198016 \pi^2
- 186624 \Sigma
+ 3079296 \zeta(3)
+ 4039632
\right. \right. \nonumber \\
&& \left. \left. ~~~~~~~~
-\, 303264 \frac{\ln^2(3) \pi}{\sqrt{3}}
+ 3639168 \frac{\ln(3) \pi}{\sqrt{3}}
+ 325728 \frac{\pi^3}{\sqrt{3}}
\right] C_F T_F \Nf \right. \nonumber \\
&& \left. ~~~~+ \left[
176256 \psi^\prime(\third)
- 117504 \pi^2 
- 3494016 
\right] T_F^2 \Nf^2 \right. \nonumber \\
&& \left. ~~~~+ \left[
138240 \psi^\prime(\third) \pi^2
- 103680 (\psi^\prime(\third))^2 
+ 1537056 \psi^\prime(\third) 
\right. \right. \nonumber \\
&& \left. \left. ~~~~~~~~
-~ 34992 \psi^{\prime\prime\prime}(\third) 
- 1679616 s_2(\pisix) 
+ 3359232 s_2(\pitwo) 
\right. \right. \nonumber \\
&& \left. \left. ~~~~~~~~
+~ 2799360 s_3(\pisix) 
- 2239488 s_3(\pitwo) 
+ 47232 \pi^4 
\right. \right. \nonumber \\
&& \left. \left. ~~~~~~~~
-~ 1024704 \pi^2
+ 139968 \Sigma
- 1399680 \zeta(3)
+ 729648
\right. \right. \nonumber \\
&& \left. \left. ~~~~~~~~
-\, 11664 \frac{\ln^2(3) \pi}{\sqrt{3}}
+ 139968 \frac{\ln(3) \pi}{\sqrt{3}}
+ 12528 \frac{\pi^3}{\sqrt{3}}
\right] C_F^2 \right] \frac{C_F a^3}{157464} \nonumber \\
&& +~ O(a^4) 
\end{eqnarray} 
and 
\begin{eqnarray}
\left. \frac{}{} \gamma^{W_3}_{33}(a,0) 
\right|_{\mbox{\footnotesize{RI$^\prime$/SMOM}}} &=&  O(a^4) ~. 
\end{eqnarray} 
Again the final expression is the same as that for the vector operator. We note
that we have not given the complete matrix at three loops. This is because the
complete three loop $\MSbar$ matrix is not known. The $12$ and $13$ elements
were not determined in \cite{12} as there was not sufficient information to 
disentangle the relation between the two counterterms. As was noted in
\cite{12} only a four loop computation could resolve this. However, the 
diagonal elements as well as the $23$ element are now available in the 
RI${}^\prime$/SMOM scheme for comparison with the $\MSbar$ expressions. In 
numerical form these anomalous dimensions are 
\begin{eqnarray}
\left. \gamma^{W_3}_{11}(a,0) \right|_{\mbox{\footnotesize{RI$^\prime$/SMOM}}} 
&=& 5.5555556 a ~+~ \left[ 161.5815415 - 10.6202306 \Nf \right] a^2 \nonumber \\
&& +~ \left[ 6275.4956875 - 993.6072317 \Nf + 24.6390623 \Nf^2 \right] 
a^3 ~+~ O(a^4) ~, \nonumber \\
\left. \gamma^{W_3}_{22}(a,0) \right|_{\mbox{\footnotesize{RI$^\prime$/SMOM}}} 
&=& 3.5555556 a ~+~ \left[ 104.7024244 - 6.5770518 \Nf \right] a^2 \nonumber \\
&& +~ \left[ 4010.9803829 - 624.8817671 \Nf + 14.9653337 \Nf^2 \right] 
a^3 ~+~ O(a^4) ~, \nonumber \\
\left. \gamma^{W_3}_{23}(a,0) \right|_{\mbox{\footnotesize{RI$^\prime$/SMOM}}} 
&=& -~ 1.7777778 a ~-~ \left[ 46.3028612 - 2.7468825 \Nf \right] a^2 
\nonumber \\
&& -~ \left[ 1692.0143513 - 265.3339715 \Nf + 6.0846171 \Nf^2 \right] a^3 ~+~ 
O(a^4) ~, \nonumber \\
\left. \gamma^{W_3}_{33}(a,0) \right|_{\mbox{\footnotesize{RI$^\prime$/SMOM}}} 
&=& O(a^4)
\end{eqnarray} 
for $SU(3)$. For comparison the corresponding $\MSbar$ values are
\begin{eqnarray}
\left. \gamma^{W_3}_{11}(a,0) \right|_{\mbox{\footnotesize{$\MSbars$}}} 
&=& 5.5555556 a ~+~ \left[ 70.8847737 - 5.1234568 \Nf \right] a^2 \nonumber \\
&& +~ \left[ 1244.9136024 - 199.6373883 \Nf - 1.7620027 \Nf^2 \right] 
a^3 ~+~ O(a^4) ~, \nonumber \\
\left. \gamma^{W_3}_{22}(a,0) \right|_{\mbox{\footnotesize{$\MSbars$}}} 
&=& 3.5555556 a ~+~ \left[ 48.3292181 - 3.1604938 \Nf \right] a^2 \nonumber \\
&& +~ \left[ 859.4478372 - 133.4381617 \Nf - 1.2290809 \Nf^2 \right] 
a^3 ~+~ O(a^4) ~, \nonumber \\
\left. \gamma^{W_3}_{23}(a,0) \right|_{\mbox{\footnotesize{$\MSbars$}}} 
&=& -~ 1.7777778 a ~-~ \left[ 24.1646091 - 1.5802469 \Nf \right] a^2 
\nonumber \\
&& -~ \left[ 429.7239186 - 66.7190809 \Nf - 0.6145405 \Nf^2 \right] a^3 ~+~ 
O(a^4) ~, \nonumber \\
\left. \gamma^{W_3}_{33}(a,0) \right|_{\mbox{\footnotesize{$\MSbars$}}} 
&=& O(a^4)
\end{eqnarray} 
which illustrates that the higher loop corrections are numerically smaller in
the $\MSbar$ scheme. 

\sect{Amplitudes.}

We devote this section to the main results which are the explicit forms of the
two loop amplitudes in both the $\MSbar$ and RI${}^\prime$/SMOM schemes. Given
the large number of amplitudes which have to be recorded we have chosen to do
this in numerical form for an arbitrary colour group. In \cite{9,11} we also
recorded the exact two loop expressions in the form of tables using the
notation 
\begin{eqnarray}
\Sigma^{{\cal O}^i}_{(j)}(p,q) &=& \left( \sum_n c^{{\cal O}^i,(1)}_{(j)\,n}
a^{(1)}_n \right) C_F a ~+~
\left( \sum_n c^{{\cal O}^i,(21)}_{(j)\,n} a^{(21)}_n \right) C_F T_F \Nf a^2
\nonumber \\
&& +~ \left( \sum_n c^{{\cal O}^i,(22)}_{(j)\,n} a^{(22)}_n \right)
C_F C_A a^2 ~+~ \left( \sum_n c^{{\cal O}^i,(23)}_{(j)\,n} a^{(23)}_n \right)
C_F^2 a^2 ~+~ O(a^3) ~. 
\end{eqnarray}
Here $a^{(l)}_n$ correspond to a basis set of numbers which naturally arise in
the computation. These include the transcendental type of numbers which have
appeared in the conversion functions and anomalous dimensions earlier. In 
addition the set $a^{(l)}_n$ also included the gauge parameter dependence. The 
labelling is chosen in such a way that $l$ indicates the loop order and in 
addition at two loops a second digit is appended to reference the colour group 
Casimir associated with that set. The other quantities, 
$c^{{\cal O}^i,(1)}_{(j)\,n}$, are the actual rational coefficients of 
interest. The main reason for recalling this notation here is that attached to 
this article is an electronic file containing all the coefficients for all the 
amplitudes in both schemes in their exact forms in terms of the basis of 
numbers $a^{(l)}_n$. Therefore this is intended to allow for comparison with 
other results such as the relation of $\partial W_3$ and $\partial \partial 
W_3$ with $W_2$ and $\partial W_2$ respectively. 

Therefore we now carry out the mundane task of recording the results. First, in
the $\MSbar$ scheme we have\footnote{Attached to this article is an electronic
file containing the explicit analytic forms for all the amplitudes presented in
this section.} 
\begin{eqnarray}
\left. \Sigma^{W_3}_{(1)}(p,q) \right|_{\MSbars} &=& -~ \left[ 0.1280942
+ 0.0401079 \alpha \right] a \nonumber \\
&& -~ \left[ 3.5739633 + 0.5909329 \alpha + 0.1445330 \alpha^2 - 0.3927663 \Nf
\right] a^2 ~+~ O(a^3) ~, \nonumber \\
\left. \Sigma^{W_3}_{(2)}(p,q) \right|_{\MSbars} &=& -~ \left[ 0.6481481
+ 0.1311651 \alpha \right] a \nonumber \\
&& -~ \left[ 11.9214676 + 0.6190048 \alpha + 0.4924713 \alpha^2 - 1.6685976 \Nf
\right] a^2 ~+~ O(a^3) ~, \nonumber \\
\left. \Sigma^{W_3}_{(3)}(p,q) \right|_{\MSbars} &=& -~ 0.3333333 ~+~ \left[ 
1.5801848 + 0.0617905 \alpha \right] a \nonumber \\
&& +~ \left[ 23.3909330 - 0.0454454 \alpha + 0.3587242 \alpha^2 - 2.8780596 \Nf
\right] a^2 ~+~ O(a^3) ~, \nonumber \\
\left. \Sigma^{W_3}_{(4)}(p,q) \right|_{\MSbars} &=& \left[ 0.3425460
+ 0.0484212 \alpha \right] a \nonumber \\
&& +~ \left[ 4.5342388 + 1.1292147 \alpha + 0.1703277 \alpha^2 - 0.2361911 \Nf
\right] a^2 ~+~ O(a^3) ~, \nonumber \\
\left. \Sigma^{W_3}_{(5)}(p,q) \right|_{\MSbars} &=& \left[ 0.4238476
+ 0.0751598 \alpha \right] a \nonumber \\
&& +~ \left[ 4.5935423 + 1.2759751 \alpha + 0.2619655 \alpha^2 - 0.3660535 \Nf
\right] a^2 ~+~ O(a^3) ~, \nonumber \\
\left. \Sigma^{W_3}_{(6)}(p,q) \right|_{\MSbars} &=& \left[ 0.5842793
+ 0.1503196 \alpha \right] a \nonumber \\
&& +~ \left[ 4.8574744 + 1.7081275 \alpha + 0.5237204 \alpha^2 - 0.6841872 \Nf
\right] a^2 ~+~ O(a^3) ~, \nonumber \\
\left. \Sigma^{W_3}_{(7)}(p,q) \right|_{\MSbars} &=& \left[ 1.0677459
+ 0.6648679 \alpha \right] a \nonumber \\
&& +~ \left[ 23.3776082 + 5.5517350 \alpha + 2.4108591 \alpha^2 - 2.1261604 \Nf
\right] a^2 ~+~ O(a^3) ~, \nonumber \\
\left. \Sigma^{W_3}_{(8)}(p,q) \right|_{\MSbars} &=& \left[ 0.2222222
+ 0.1553757 \alpha \right] a \nonumber \\
&& +~ \left[ 7.1027076 + 2.4806545 \alpha + 0.5427210 \alpha^2 - 0.1633838 \Nf
\right] a^2 ~+~ O(a^3) ~, \nonumber \\
\left. \Sigma^{W_3}_{(9)}(p,q) \right|_{\MSbars} &=& \left[ 0.2829270
+ 0.2088529 \alpha \right] a \nonumber \\
&& +~ \left[ 9.2546037 + 3.0300483 \alpha + 0.7383424 \alpha^2 - 0.2645584 \Nf
\right] a^2 ~+~ O(a^3) ~, \nonumber \\
\left. \Sigma^{W_3}_{(10)}(p,q) \right|_{\MSbars} &=& \left[ 0.4773248
+ 0.3374900 \alpha \right] a \nonumber \\
&& +~ \left[ 15.4402123 + 3.8373519 \alpha + 1.2238427 \alpha^2 - 0.7240161 \Nf
\right] a^2 ~+~ O(a^3) ~, \nonumber \\
\left. \Sigma^{W_3}_{(11)}(p,q) \right|_{\MSbars} &=& \left[ 0.5615110
+ 0.1503196 \alpha \right] a \nonumber \\
&& +~ \left[ 7.6079600 + 1.7765023 \alpha + 0.5439553 \alpha^2 - 0.6061888 \Nf
\right] a^2 ~+~ O(a^3) ~, \nonumber \\
\left. \Sigma^{W_3}_{(12)}(p,q) \right|_{\MSbars} &=& -~ 0.0607048 a 
\nonumber \\
&& -~ \left[ 1.2759196 + 0.2443179 \alpha + 0.0050587 \alpha^2 - 0.0499049 \Nf
\right] a^2 ~+~ O(a^3) ~, \nonumber \\
\left. \Sigma^{W_3}_{(13)}(p,q) \right|_{\MSbars} &=& -~ 0.1214095 a 
\nonumber \\
&& -~ \left[ 2.5537039 + 0.3188150 \alpha + 0.0101175 \alpha^2 - 0.1270706 \Nf
\right] a^2 ~+~ O(a^3) ~, \nonumber \\
\left. \Sigma^{W_3}_{(14)}(p,q) \right|_{\MSbars} &=& -~ 0.4588995 a 
\nonumber \\
&& -~ \left[ 8.6460262 - 0.5239987 \alpha + 0.0382416 \alpha^2 - 0.7578066 \Nf
\right] a^2 \nonumber \\
&& +~ O(a^3) ~, 
\end{eqnarray} 
\begin{eqnarray}
\left. \Sigma^{\partial W_3}_{(1)}(p,q) \right|_{\MSbars} &=& 
-~ \left[ 0.5833178 + 0.1481415 \alpha \right] a \nonumber \\
&& -~ \left[ 12.4616428 + 0.8098939 \alpha + 0.5536285 \alpha^2 - 1.6747894 \Nf
\right] a^2 ~+~ O(a^3) ~, \nonumber \\
\left. \Sigma^{\partial W_3}_{(2)}(p,q) \right|_{\MSbars} &=& -~ 0.1666667 ~+~
\left[ 0.2708217 - 0.0971989 \alpha \right] a \nonumber \\
&& +~ \left[ 1.0208053 - 0.5371502 \alpha - 0.3019999 \alpha^2 + 0.0393764 \Nf
\right] a^2 ~+~ O(a^3) ~, \nonumber \\
\left. \Sigma^{\partial W_3}_{(3)}(p,q) \right|_{\MSbars} &=& -~ 0.3333333 ~+~
\left[ 1.1249617 - 0.0462497 \alpha \right] a \nonumber \\
&& +~ \left[ 14.5032535 - 0.2644064 \alpha - 0.0503712 \alpha^2 - 1.5960365 \Nf
\right] a^2 ~+~ O(a^3) ~, \nonumber \\
\left. \Sigma^{\partial W_3}_{(4)}(p,q) \right|_{\MSbars} &=& 
\left[ 0.6265587 + 0.1018984 \alpha \right] a \nonumber \\
&& +~ \left[ 7.8618848 + 1.9742471 \alpha + 0.3608903 \alpha^2 - 0.4799771 \Nf
\right] a^2 ~+~ O(a^3) ~, \nonumber \\
\left. \Sigma^{\partial W_3}_{(5)}(p,q) \right|_{\MSbars} &=& 
\left[ 0.7602518 + 0.1698307 \alpha \right] a \nonumber \\
&& +~ \left[ 8.2392595 + 2.2694139 \alpha + 0.5976297 \alpha^2 - 0.7417023 \Nf
\right] a^2 ~+~ O(a^3) ~, \nonumber \\
\left. \Sigma^{\partial W_3}_{(6)}(p,q) \right|_{\MSbars} &=& 
\left[ 0.9886158 + 0.4198773 \alpha \right] a \nonumber \\
&& +~ \left[ 15.7278789 + 4.1413534 \alpha + 1.5021215 \alpha^2 - 1.3862056 \Nf
\right] a^2 ~+~ O(a^3) ~, \nonumber \\
\left. \Sigma^{\partial W_3}_{(7)}(p,q) \right|_{\MSbars} &=& 
\left[ 1.3116507 + 0.8520382 \alpha \right] a \nonumber \\
&& +~ \left[ 30.3277430 + 7.5900655 \alpha + 3.0743658 \alpha^2 - 2.4134871 \Nf
\right] a^2 ~+~ O(a^3) ~, \nonumber \\
\left. \Sigma^{\partial W_3}_{(8)}(p,q) \right|_{\MSbars} &=& 
\left[ 0.4661270 + 0.3425460 \alpha \right] a \nonumber \\
&& +~ \left[ 14.0528425 + 4.5189850 \alpha + 1.2062277 \alpha^2 - 0.4507105 \Nf
\right] a^2 ~+~ O(a^3) ~, \nonumber \\
\left. \Sigma^{\partial W_3}_{(9)}(p,q) \right|_{\MSbars} &=& 
\left[ 0.6872635 + 0.4784106 \alpha \right] a \nonumber \\
&& +~ \left[ 20.1250083 + 5.4632742 \alpha + 1.7167435 \alpha^2 - 0.9665767 \Nf
\right] a^2 ~+~ O(a^3) ~, \nonumber \\
\left. \Sigma^{\partial W_3}_{(10)}(p,q) \right|_{\MSbars} &=& 
\left[ 0.8137291 + 0.4321609 \alpha \right] a \nonumber \\
&& +~ \left[ 19.0859295 + 4.8307907 \alpha + 1.5595069 \alpha^2 - 1.0996648 \Nf
\right] a^2 ~+~ O(a^3) ~, \nonumber \\
\left. \Sigma^{\partial W_3}_{(11)}(p,q) \right|_{\MSbars} &=& 
\left[ 0.8455237 + 0.2037969 \alpha \right] a \nonumber \\
&& +~ \left[ 10.9356060 + 2.6215347 \alpha + 0.7345179 \alpha^2 - 0.8499747 \Nf
\right] a^2 ~+~ O(a^3) ~, \nonumber \\
\left. \Sigma^{\partial W_3}_{(12)}(p,q) \right|_{\MSbars} &=& 
-~ 0.1481481 a \nonumber \\
&& -~ \left[ 2.9461750 + 0.4422360 \alpha + 0.0123457 \alpha^2 - 0.1476260 \Nf
\right] a^2 ~+~ O(a^3) ~, \nonumber \\
\left. \Sigma^{\partial W_3}_{(13)}(p,q) \right|_{\MSbars} &=& 
-~ 0.3472455 a \nonumber \\
&& -~ \left[ 6.6312283 + 0.0580777 \alpha + 0.0289371 \alpha^2 - 0.5015769 \Nf
\right] a^2 ~+~ O(a^3) ~, \nonumber \\
\left. \Sigma^{\partial W_3}_{(14)}(p,q) \right|_{\MSbars} &=& 
-~ 0.5463429 a \nonumber \\
&& -~ \left[ 10.3162815 - 0.3260806 \alpha + 0.0455286 \alpha^2 - 0.8555277 \Nf
\right] a^2 \nonumber \\
&& +~ O(a^3) 
\end{eqnarray}
and 
\begin{eqnarray}
\left. \Sigma^{\partial\partial W_3}_{(1)}(p,q) \right|_{\MSbars} &=& 
\left. \Sigma^{\partial\partial W_3}_{(2)}(p,q) \right|_{\MSbars} ~=~ 
\left. \Sigma^{\partial\partial W_3}_{(3)}(p,q) \right|_{\MSbars} \nonumber \\ 
&=& -~ 0.3333333 ~+~ \left[ 0.5416434 - 0.1943979 \alpha \right] a \nonumber \\
&& +~ \left[ 2.0416107 - 1.0743003 \alpha - 0.6039997 \alpha^2 + 0.0787529 \Nf
\right] a^2 \nonumber \\
&& +~ O(a^3) ~, \nonumber \\
\left. \Sigma^{\partial\partial W_3}_{(4)}(p,q) \right|_{\MSbars} &=&  
\left. \Sigma^{\partial\partial W_3}_{(11)}(p,q) \right|_{\MSbars} \nonumber \\ 
&=& \left[ 1.4720825 + 0.3056953 \alpha \right] a \nonumber \\
&& +~ \left[ 18.7974908 + 4.5957818 \alpha + 1.0954083 \alpha^2 - 1.3299518 \Nf
\right] a^2 \nonumber \\
&& +~ O(a^3) ~, \nonumber \\
\left. \Sigma^{\partial\partial W_3}_{(5)}(p,q) \right|_{\MSbars} &=&  
\left. \Sigma^{\partial\partial W_3}_{(10)}(p,q) \right|_{\MSbars} \nonumber \\ 
&=& \left[ 1.5739809 + 0.6019916 \alpha \right] a \nonumber \\
&& +~ \left[ 27.3251890 + 7.1002047 \alpha + 2.1571366 \alpha^2 - 1.8413671 \Nf
\right] a^2 \nonumber \\
&& +~ O(a^3) ~, \nonumber \\
\left. \Sigma^{\partial\partial W_3}_{(6)}(p,q) \right|_{\MSbars} &=&  
\left. \Sigma^{\partial\partial W_3}_{(9)}(p,q) \right|_{\MSbars} \nonumber \\ 
&=& \left[ 1.6758793 + 0.8982879 \alpha \right] a \nonumber \\
&& +~ \left[ 35.8528872 + 9.6046276 \alpha + 3.2188650 \alpha^2 - 2.3527823 \Nf
\right] a^2 \nonumber \\
&& +~ O(a^3) ~, \nonumber \\
\left. \Sigma^{\partial\partial W_3}_{(7)}(p,q) \right|_{\MSbars} &=& 
\left. \Sigma^{\partial\partial W_3}_{(8)}(p,q) \right|_{\MSbars} \nonumber \\ 
&=& \left[ 1.7777778 + 1.1945842 \alpha \right] a \nonumber \\
&& +\, \left[ 44.3805855 + 12.1090504 \alpha + 4.2805934 \alpha^2 - 2.8641975 
\Nf \right] a^2 \nonumber \\
&& +~ O(a^3) ~, \nonumber \\
\left. \Sigma^{\partial\partial W_3}_{(12)}(p,q) \right|_{\MSbars} &=& 
\left. \Sigma^{\partial\partial W_3}_{(13)}(p,q) \right|_{\MSbars} ~=~ 
\left. \Sigma^{\partial\partial W_3}_{(14)}(p,q) \right|_{\MSbars} \nonumber \\ 
&=& -~ 0.6944910 a \nonumber \\
&& -\, \left[ 13.2624565 + 0.1161554 \alpha + 0.0578743 \alpha^2 - 1.0031537 
\Nf \right] a^2 \nonumber \\
&& +~ O(a^3) ~. 
\end{eqnarray}
With the two loop corrections computed we note that the results for some of the
$\partial W_3$ and $\partial\partial W_3$ amplitudes are proportional to those
of lower moment operators. These relations were noted at one loop in \cite{10}
and extend to two loops now. For instance, channels $4$ and $7$ of 
$\partial\partial W_3$ are related to channels $3$ and $5$ respectively of
$\partial W_2$, \cite{9,10}, or equally channels $2$ and $3$ of the vector
current, \cite{10,11}. Also channels $4$ and $7$ of $\partial W_3$ are 
proportional to channels $5$ and $7$ respectively of $W_2$. That not all 
channels have a similar relation is due to the imbalance of indices of the 
operator inserted in the Green's function. Also the absence of direct equality 
stems from the difference in the two bases used for each operator moment. 
Though this underlying agreement is a partial check on our computation. Equally
the equality of several of the $\partial\partial W_3$ channels with themselves
reflects the symmetric nature of this particular operator and is another minor
calculational check.  
 
The expressions for the RI${}^\prime$/SMOM scheme amplitudes have a similar
structure to the $\MSbar$ ones. However, given the nature of the scheme
definition which we have used then there are no corrections to channels $1$ and
$2$ of $W_3$ as well as $1$ of $\partial W_3$. For channels $3$ of $W_3$ and 
$2$ and $3$ of $\partial W_3$ only the tree term is present which we record 
with the rest of the amplitudes. We have  
\begin{eqnarray}
\Sigma^{W_3}_{(3)}(p,q) &=& -~ 0.3333333 ~+~ O(a^3) ~, \nonumber \\
\Sigma^{W_3}_{(4)}(p,q) &=& \left[ 0.3425460 + 0.0484212 \alpha \right] a 
\nonumber \\
&& +~ \left[ 4.8378239 + 1.2568504 \alpha + 0.1917623 \alpha^2
+ 0.0363159 \alpha^3 \right. \nonumber \\
&& \left. ~~~~-~ ( 0.2361911 + 0.0538013 \alpha ) \Nf \right] a^2 ~+~ O(a^3) ~, 
\nonumber \\
\Sigma^{W_3}_{(5)}(p,q) &=& \left[ 0.4238476 + 0.0751598 \alpha \right] a 
\nonumber \\
&& +~ \left[ 5.1111487 + 1.4983490 \alpha + 0.2735246 \alpha^2
+ 0.0563697 \alpha^3 \right. \nonumber \\
&& \left. ~~~~-~ ( 0.3660535 + 0.0835109 \alpha ) \Nf \right] a^2 ~+~ O(a^3) ~, 
\nonumber \\
\Sigma^{W_3}_{(6)}(p,q) &=& \left[ 0.5842793 + 0.1503196 \alpha \right] a 
\nonumber \\
&& +~ \left[ 5.9461385 + 2.2633283 \alpha + 0.5397956 \alpha^2
+ 0.1127397 \alpha^3 \right. \nonumber \\
&& \left. ~~~~-~ ( 0.6841872 + 0.1670218 \alpha ) \Nf \right] a^2 ~+~ O(a^3) ~, 
\nonumber \\
\Sigma^{W_3}_{(7)}(p,q) &=& \left[ 1.0677459 + 0.6648679 \alpha \right] a 
\nonumber \\
&& +~ \left[ 27.4054010 + 13.2965737 \alpha + 3.3654119 \alpha^2
+ 0.4986509 \alpha^3 \right. \nonumber \\
&& \left. ~~~~-~ ( 2.1261604 + 0.7387421 \alpha ) \Nf \right] a^2 ~+~ O(a^3) ~, 
\nonumber \\
\Sigma^{W_3}_{(8)}(p,q) &=& \left[ 0.2222222 + 0.1553757 \alpha \right] a 
\nonumber \\
&& +~ \left[ 6.7973293 + 3.2105607 \alpha + 0.5772861 \alpha^2
+ 0.1165318 \alpha^3 \right. \nonumber \\
&& \left. ~~~~-~ ( 0.1633838 + 0.1726396 \alpha ) \Nf \right] a^2 ~+~ O(a^3) ~, 
\nonumber \\
\Sigma^{W_3}_{(9)}(p,q) &=& \left[ 0.2829270 + 0.2088529 \alpha \right] a 
\nonumber \\
&& +~ \left[ 8.7988864 + 4.2662444 \alpha + 0.8601108 \alpha^2
+ 0.1566397 \alpha^3 \right. \nonumber \\
&& \left. ~~~~-~ ( 0.2645584 + 0.2320588 \alpha ) \Nf \right] a^2 ~+~ O(a^3) ~, 
\nonumber \\
\Sigma^{W_3}_{(10)}(p,q) &=& \left[ 0.4773248 + 0.3374900 \alpha \right] a 
\nonumber \\
&& +~ \left[ 16.2318808 + 7.5409748 \alpha + 1.7090901 \alpha^2
+ 0.2531175 \alpha^3 \right. \nonumber \\
&& \left. ~~~~-~ ( 0.7240161 + 0.3749886 \alpha ) \Nf \right] a^2 ~+~ O(a^3) ~, 
\nonumber \\
\Sigma^{W_3}_{(11)}(p,q) &=& \left[ 0.5615110 + 0.1503196 \alpha \right] a 
\nonumber \\
&& +~ \left[ 9.0337049 + 3.3169665 \alpha + 0.7493875 \alpha^2
+ 0.1127397 \alpha^3 \right. \nonumber \\
&& \left. ~~~~-~ ( 0.6061888 + 0.1670218 \alpha ) \Nf \right] a^2 ~+~ O(a^3) ~, 
\nonumber \\
\Sigma^{W_3}_{(12)}(p,q) &=& -~ 0.0607048 a \nonumber \\
&& -~ \left[ 1.0472892 + 0.1315372 \alpha + 0.0050587 \alpha^2 - 0.0499049 \Nf 
\right] a^2 ~+~ O(a^3) ~, \nonumber \\
\Sigma^{W_3}_{(13)}(p,q) &=& -~ 0.1214095 a \nonumber \\
&& -~ \left[ 2.2043458 + 0.1489816 \alpha + 0.0101175 \alpha^2 - 0.1270706 \Nf 
\right] a^2 ~+~ O(a^3) ~, \nonumber \\
\Sigma^{W_3}_{(14)}(p,q) &=& -~ 0.4588995 a \nonumber \\
&& -~ \left[ 10.4580788 - 0.5150531 \alpha + 0.0382416 \alpha^2 - 0.7578066 \Nf 
\right] a^2 \nonumber \\
&& +~ O(a^3) ~, 
\end{eqnarray}
\begin{eqnarray}
\Sigma^{\partial W_3}_{(2)}(p,q) &=& -~ 0.1666667 ~+~ O(a^3) ~~~,~~~
\Sigma^{\partial W_3}_{(3)}(p,q) ~=~ -~ 0.3333333 ~+~ O(a^3) \nonumber \\
\Sigma^{\partial W_3}_{(4)}(p,q) &=& \left[ 0.6265587 + 0.1018984 \alpha 
\right] a \nonumber \\
&& +~ \left[ 8.4968205 + 2.3224637 \alpha + 0.4090233 \alpha^2
+ 0.0764238 \alpha^3 \right. \nonumber \\
&& \left. ~~~~-~ ( 0.4799771 + 0.1132205 \alpha ) \Nf \right] a^2 ~+~ O(a^3) ~, 
\nonumber \\
\Sigma^{\partial W_3}_{(5)}(p,q) &=& \left[ 0.7602518 + 0.1698307 \alpha 
\right] a \nonumber \\
&& +~ \left[ 9.3810330 + 2.9919683 \alpha + 0.6367405 \alpha^2
+ 0.1273731 \alpha^3 \right. \nonumber \\
&& \left. ~~~~-~ ( 0.7417023 + 0.1887008 \alpha ) \Nf \right] a^2 ~+~ O(a^3) ~, 
\nonumber \\
\Sigma^{\partial W_3}_{(6)}(p,q) &=& \left[ 0.9886158 + 0.4198773 \alpha 
\right] a \nonumber \\
&& +~ \left[ 17.8616631 + 7.6725826 \alpha + 1.8610530 \alpha^2
+ 0.3149080 \alpha^3 \right. \nonumber \\
&& \left. ~~~~-~ ( 1.3862056 + 0.4665304 \alpha ) \Nf \right] a^2 ~+~ O(a^3) ~, 
\nonumber \\
\Sigma^{\partial W_3}_{(7)}(p,q) &=& \left[ 1.3116507 + 0.8520382 \alpha 
\right] a \nonumber \\
&& +~ \left[ 33.9387109 + 16.3643066 \alpha + 4.0819609 \alpha^2
+ 0.6390287 \alpha^3 \right. \nonumber \\
&& \left. ~~~~-~ ( 2.4134871 + 0.9467091 \alpha ) \Nf \right] a^2 ~+~ O(a^3) ~, 
\nonumber \\
\Sigma^{\partial W_3}_{(8)}(p,q) &=& \left[ 0.4661270 + 0.3425460 \alpha 
\right] a \nonumber \\
&& +~ \left[ 13.3306392 + 6.3052936 \alpha + 1.2938351 \alpha^2
+ 0.2569095 \alpha^3 \right. \nonumber \\
&& \left. ~~~~-~ ( 0.4507105 + 0.3806067 \alpha ) \Nf \right] a^2 ~+~ O(a^3) ~, 
\nonumber \\
\Sigma^{\partial W_3}_{(9)}(p,q) &=& \left[ 0.6872635 + 0.4784106 \alpha 
\right] a \nonumber \\
&& +~ \left[ 20.7144109 + 9.6754987 \alpha + 2.1813682 \alpha^2
+ 0.3588079 \alpha^3 \right. \nonumber \\
&& \left. ~~~~-~ ( 0.9665767 + 0.5315673 \alpha ) \Nf \right] a^2 ~+~ O(a^3) ~, 
\nonumber \\
\Sigma^{\partial W_3}_{(10)}(p,q) &=& \left[ 0.8137291 + 0.4321609 \alpha 
\right] a \nonumber \\
&& +~ \left[ 20.5017650 + 9.0345941 \alpha + 2.0723060 \alpha^2
+ 0.3241207 \alpha^3 \right. \nonumber \\
&& \left. ~~~~-~ ( 1.0996648 + 0.4801788 \alpha ) \Nf \right] a^2 ~+~ O(a^3) ~, 
\nonumber \\
\Sigma^{\partial W_3}_{(11)}(p,q) &=& \left[ 0.8455237 + 0.2037969 \alpha 
\right] a \nonumber \\
&& +~ \left[ 12.6927015 + 4.3825798 \alpha + 0.9666484 \alpha^2
+ 0.1528477 \alpha^3 \right. \nonumber \\
&& \left. ~~~~-~ ( 0.8499747 + 0.2264410 \alpha ) \Nf \right] a^2 ~+~ O(a^3) ~, 
\nonumber \\
\Sigma^{\partial W_3}_{(12)}(p,q) &=& -~ 0.1481481 a \nonumber \\
&& -~ \left[ 2.4900831 + 0.1788615 \alpha + 0.0123457 \alpha^2 - 0.1476260 \Nf 
\right] a^2 ~+~ O(a^3) ~, \nonumber \\
\Sigma^{\partial W_3}_{(13)}(p,q) &=& -~ 0.3472455 a \nonumber \\
&& -~ \left[ 7.1954780 - 0.1444336 \alpha + 0.0289371 \alpha^2 - 0.5015769 \Nf 
\right] a^2 ~+~ O(a^3) ~, \nonumber \\
\Sigma^{\partial W_3}_{(14)}(p,q) &=& -~ 0.5463429 a \nonumber \\
&& -~ \left[ 11.9008728 - 0.4677288 \alpha + 0.0455286 \alpha^2 - 0.8555277 \Nf 
\right] a^2 \nonumber \\
&& +~ O(a^3) 
\end{eqnarray}
and 
\begin{eqnarray}
\Sigma^{\partial\partial W_3}_{(1)}(p,q) &=& 
\Sigma^{\partial\partial W_3}_{(2)}(p,q) ~=~ 
\Sigma^{\partial\partial W_3}_{(3)}(p,q) \nonumber \\ 
&=& -~ 0.3333333 ~+~ \left[ 0.5416433 - 0.2500466 \alpha \right] a \nonumber \\
&& +~ \left[ 10.5296794 + 4.0831575 \alpha + 0.9376747 \alpha^2 
+ 0.1875349 \alpha^3 \right. \nonumber \\
&& \left. ~~~~-~ ( 0.6990249 + 0.2778295 \alpha ) \Nf \right] a^2 ~+~ O(a^3) ~, 
\nonumber \\
\Sigma^{\partial\partial W_3}_{(4)}(p,q) &=& 
\Sigma^{\partial\partial W_3}_{(11)}(p,q) \nonumber \\ 
&=& \left[ 1.4720825 + 0.3056953 \alpha \right] a \nonumber \\
&& +~ \left[ 18.7974908 + 5.1040424 \alpha + 1.1463575 \alpha^2 
+ 0.2292715 \alpha^3 \right. \nonumber \\
&& \left. ~~~~-~ ( 1.3299518 + 0.3396615 \alpha ) \Nf \right] a^2 ~+~ O(a^3) ~, 
\nonumber \\
\Sigma^{\partial\partial W_3}_{(5)}(p,q) &=& 
\Sigma^{\partial\partial W_3}_{(10)}(p,q) \nonumber \\ 
&=& \left[ 1.5739809 + 0.6019916 \alpha \right] a \nonumber \\
&& +~ \left[ 27.3251890 + 9.8676624 \alpha + 2.2574686 \alpha^2 
+ 0.4514937 \alpha^3 \right. \nonumber \\
&& \left. ~~~~-~ ( 1.8413671 + 0.6688796 \alpha ) \Nf \right] a^2 ~+~ O(a^3) ~, 
\nonumber \\
\Sigma^{\partial\partial W_3}_{(6)}(p,q) &=& 
\Sigma^{\partial\partial W_3}_{(9)}(p,q) \nonumber \\ 
&=& \left[ 1.6758793 + 0.8982879 \alpha \right] a \nonumber \\
&& +~ \left[ 35.8528873 + 14.6312824 \alpha + 3.3685797 \alpha^2 
+ 0.6737159 \alpha^3 \right. \nonumber \\
&& \left. ~~~~-~ ( 2.3527823 + 0.9980977 \alpha ) \Nf \right] a^2 ~+~ O(a^3) ~, 
\nonumber \\
\Sigma^{\partial\partial W_3}_{(7)}(p,q) &=& 
\Sigma^{\partial\partial W_3}_{(8)}(p,q) \nonumber \\ 
&=& \left[ 1.7777778 + 1.1945842 \alpha \right] a \nonumber \\
&& +~ \left[ 44.3805855 + 19.3949025 \alpha + 4.4796908 \alpha^2 
+ 0.8959382 \alpha^3 \right. \nonumber \\
&& \left. ~~~~-~ ( 2.8641975 + 1.3272318 \alpha ) \Nf \right] a^2 ~+~ O(a^3) ~, 
\nonumber \\
\Sigma^{\partial\partial W_3}_{(12)}(p,q) &=& 
\Sigma^{\partial\partial W_3}_{(13)}(p,q) ~=~
\Sigma^{\partial\partial W_3}_{(14)}(p,q)  \nonumber \\ 
&=& -~ 0.6944910 a \nonumber \\
&& -~ \left[ 13.2624565 - 0.8098326 \alpha + 0.0578743 \alpha^2 - 1.0031537 \Nf
\right] a^2 \nonumber \\
&& +~ O(a^3) ~. 
\end{eqnarray}
The same relations between the various channels noted earlier for the $\MSbar$
scheme apply to the corresponding amplitudes in the RI${}^\prime$/SMOM scheme.

\sect{Discussion.}

We conclude our discussions with brief remarks. Clearly we have provided the
full two loop structure of the Green's function with level $W_3$ operators
inserted in a quark $2$-point function in two renormalization schemes. The
underlying renormalization allowed for the computation of the conversion
functions from the RI${}^\prime$/SMOM scheme to $\MSbar$ and we have come to 
the same conclusion as \cite{9,10} that the series convergence is marginally 
worse than the conversion functions of the RI${}^\prime$ scheme which has
potential infrared issues. Though as we have argued that in some sense 
RI${}^\prime$ is not as full a scheme as RI${}^\prime$/SMOM as there is no 
access to off-diagonal elements of the mixing matrix. Although one possibility 
of improving the convergence rests in the redefinition of the 
RI${}^\prime$/SMOM scheme to take account of the structure of more amplitudes. 
This could be achieved by a different choice of tensor basis for the Green's 
function. However, in providing the full structure of the Green's function at 
the symmetric point in $\MSbar$ one is free to perform the renormalization in 
any scheme of their choosing before converting to $\MSbar$ as the reference 
scheme. Therefore, our results are reasonably comprehensive so as to allow 
others to be flexible in how they choose to define their own version of 
RI${}^\prime$/SMOM.

\vspace{1cm}
\noindent
{\bf Acknowledgement.} The author thanks Dr. P.E.L. Rakow for useful 
discussions.

\appendix

\sect{Basis tensors and projection matrix.} 

The explicit forms of the fourteen basis tensors in $d$-dimensions are 
\begin{eqnarray}
{\cal P}^{W_3}_{(1) \mu \nu \sigma}(p,q) &=& \frac{1}{\mu^2} \left[
\gamma_\mu p_\nu p_\sigma + \gamma_\nu p_\sigma p_\mu
+ \gamma_\sigma p_\mu p_\nu \right] ~+~ \frac{1}{[d+2]} \left[ 
\gamma_\mu \eta_{\nu\sigma} + \gamma_\nu \eta_{\sigma\mu}
+ \gamma_\sigma \eta_{\mu\nu} \right] \nonumber \\
&& -~ \frac{2\pslash}{[d+2]\mu^2} \left[ \eta_{\mu\nu} p_\sigma
+ \eta_{\nu\sigma} p_\mu + \eta_{\sigma\mu} p_\nu \right] ~, \nonumber \\ 
{\cal P}^{W_3}_{(2) \mu \nu \sigma}(p,q) &=& \frac{1}{\mu^2} \left[
\gamma_\mu p_\nu q_\sigma + \gamma_\nu p_\sigma q_\mu
+ \gamma_\sigma p_\mu q_\nu + \gamma_\mu q_\nu p_\sigma 
+ \gamma_\nu q_\sigma p_\mu + \gamma_\sigma q_\mu p_\nu \right] \nonumber \\
&& -~ \frac{1}{[d+2]} \left[ 
\gamma_\mu \eta_{\nu\sigma} + \gamma_\nu \eta_{\sigma\mu}
+ \gamma_\sigma \eta_{\mu\nu} \right] ~-~ \frac{2\pslash}{[d+2]\mu^2} \left[ 
\eta_{\mu\nu} q_\sigma + \eta_{\nu\sigma} q_\mu + \eta_{\sigma\mu} q_\nu 
\right] \nonumber \\ 
&& -~ \frac{2\qslash}{[d+2]\mu^2} \left[ \eta_{\mu\nu} p_\sigma
+ \eta_{\nu\sigma} p_\mu + \eta_{\sigma\mu} p_\nu \right] ~, \nonumber \\ 
{\cal P}^{W_3}_{(3) \mu \nu \sigma}(p,q) &=& \frac{1}{\mu^2} \left[
\gamma_\mu q_\nu q_\sigma + \gamma_\nu q_\sigma q_\mu
+ \gamma_\sigma q_\mu q_\nu \right] ~+~ \frac{1}{[d+2]} \left[ 
\gamma_\mu \eta_{\nu\sigma} + \gamma_\nu \eta_{\sigma\mu}
+ \gamma_\sigma \eta_{\mu\nu} \right] \nonumber \\
&& -~ \frac{2\qslash}{[d+2]\mu^2} \left[ \eta_{\mu\nu} q_\sigma
+ \eta_{\nu\sigma} q_\mu + \eta_{\sigma\mu} q_\nu \right] ~, \nonumber \\ 
{\cal P}^{W_3}_{(4) \mu \nu \sigma}(p,q) &=& \frac{\pslash}{\mu^4} 
p_\mu p_\nu p_\sigma ~+~ \frac{\pslash}{[d+2]\mu^2} \left[ 
\eta_{\mu\nu} p_\sigma + \eta_{\nu\sigma} p_\mu + \eta_{\sigma\mu} p_\nu 
\right] ~, \nonumber \\
{\cal P}^{W_3}_{(5) \mu \nu \sigma}(p,q) &=& \frac{\pslash}{\mu^4} \left[ 
p_\mu p_\nu q_\sigma + p_\mu q_\nu p_\sigma + q_\mu p_\nu p_\sigma \right] 
\nonumber \\
&& -~ \frac{\pslash}{[d+2]\mu^2} \left[ \eta_{\mu\nu} p_\sigma 
- \eta_{\mu\nu} q_\sigma + \eta_{\nu\sigma} p_\mu - \eta_{\nu\sigma} q_\mu 
+ \eta_{\sigma\mu} p_\nu - \eta_{\sigma\mu} q_\nu \right] ~, \nonumber \\ 
{\cal P}^{W_3}_{(6) \mu \nu \sigma}(p,q) &=& \frac{\pslash}{\mu^4} \left[ 
p_\mu q_\nu q_\sigma + q_\mu p_\nu q_\sigma + q_\mu q_\nu p_\sigma \right] 
\nonumber \\
&& +~ \frac{\pslash}{[d+2]\mu^2} \left[ \eta_{\mu\nu} p_\sigma 
- \eta_{\mu\nu} q_\sigma + \eta_{\nu\sigma} p_\mu - \eta_{\nu\sigma} q_\mu 
+ \eta_{\sigma\mu} p_\nu - \eta_{\sigma\mu} q_\nu \right] ~, \nonumber \\ 
{\cal P}^{W_3}_{(7) \mu \nu \sigma}(p,q) &=& \frac{\pslash}{\mu^4} 
q_\mu q_\nu q_\sigma ~+~ \frac{\pslash}{[d+2]\mu^2} \left[ 
\eta_{\mu\nu} q_\sigma + \eta_{\nu\sigma} q_\mu + \eta_{\sigma\mu} q_\nu 
\right] ~, \nonumber \\
{\cal P}^{W_3}_{(8) \mu \nu \sigma}(p,q) &=& \frac{\qslash}{\mu^4} 
p_\mu p_\nu p_\sigma ~+~ \frac{\qslash}{[d+2]\mu^2} \left[ 
\eta_{\mu\nu} p_\sigma + \eta_{\nu\sigma} p_\mu + \eta_{\sigma\mu} p_\nu 
\right] ~, \nonumber \\
{\cal P}^{W_3}_{(9) \mu \nu \sigma}(p,q) &=& \frac{\qslash}{\mu^4} \left[ 
p_\mu p_\nu q_\sigma + p_\mu q_\nu p_\sigma + q_\mu p_\nu p_\sigma \right] 
\nonumber \\
&& -~ \frac{\qslash}{[d+2]\mu^2} \left[ \eta_{\mu\nu} p_\sigma 
- \eta_{\mu\nu} q_\sigma + \eta_{\nu\sigma} p_\mu - \eta_{\nu\sigma} q_\mu 
+ \eta_{\sigma\mu} p_\nu - \eta_{\sigma\mu} q_\nu \right] ~, \nonumber \\ 
{\cal P}^{W_3}_{(10) \mu \nu \sigma}(p,q) &=& \frac{\qslash}{\mu^4} \left[ 
p_\mu q_\nu q_\sigma + q_\mu p_\nu q_\sigma + q_\mu q_\nu p_\sigma \right] 
\nonumber \\
&& +~ \frac{\qslash}{[d+2]\mu^2} \left[ \eta_{\mu\nu} p_\sigma 
- \eta_{\mu\nu} q_\sigma + \eta_{\nu\sigma} p_\mu - \eta_{\nu\sigma} q_\mu 
+ \eta_{\sigma\mu} p_\nu - \eta_{\sigma\mu} q_\nu \right] ~, \nonumber \\ 
{\cal P}^{W_3}_{(11) \mu \nu \sigma}(p,q) &=& \frac{\qslash}{\mu^4} 
q_\mu q_\nu q_\sigma ~+~ \frac{\qslash}{[d+2]\mu^2} \left[ 
\eta_{\mu\nu} q_\sigma + \eta_{\nu\sigma} q_\mu + \eta_{\sigma\mu} q_\nu 
\right] ~, \nonumber \\
{\cal P}^{W_3}_{(12) \mu \nu \sigma}(p,q) &=& \frac{1}{\mu^4} \left[ 
\Gamma_{(3) \, \mu p q} p_\nu p_\sigma +
\Gamma_{(3) \, \nu p q} p_\mu p_\sigma +
\Gamma_{(3) \, \sigma p q} p_\mu p_\nu \right] \nonumber \\
&& +~ \frac{1}{[d+2]\mu^2} \left[ 
\Gamma_{(3) \, \mu p q} \eta_{\nu\sigma} +
\Gamma_{(3) \, \nu p q} \eta_{\mu\sigma} +
\Gamma_{(3) \, \sigma p q} \eta_{\mu\nu} \right] ~, \nonumber \\ 
{\cal P}^{W_3}_{(13) \mu \nu \sigma}(p,q) &=& \frac{1}{\mu^4} \left[ 
\Gamma_{(3) \, \mu p q} p_\nu q_\sigma +
\Gamma_{(3) \, \nu p q} p_\mu q_\sigma +
\Gamma_{(3) \, \sigma p q} p_\mu q_\nu \right. \nonumber \\
&& \left. ~~~~+~ \Gamma_{(3) \, \mu p q} q_\nu p_\sigma +
\Gamma_{(3) \, \nu p q} q_\mu p_\sigma +
\Gamma_{(3) \, \sigma p q} q_\mu p_\nu \right] \nonumber \\
&& -~ \frac{1}{[d+2]\mu^2} \left[ 
\Gamma_{(3) \, \mu p q} \eta_{\nu\sigma} +
\Gamma_{(3) \, \nu p q} \eta_{\mu\sigma} +
\Gamma_{(3) \, \sigma p q} \eta_{\mu\nu} \right] ~, \nonumber \\ 
{\cal P}^{W_3}_{(14) \mu \nu \sigma}(p,q) &=& \frac{1}{\mu^4} \left[ 
\Gamma_{(3) \, \mu p q} q_\nu q_\sigma +
\Gamma_{(3) \, \nu p q} q_\mu q_\sigma +
\Gamma_{(3) \, \sigma p q} q_\mu q_\nu \right] \nonumber \\
&& +~ \frac{1}{[d+2]\mu^2} \left[ 
\Gamma_{(3) \, \mu p q} \eta_{\nu\sigma} +
\Gamma_{(3) \, \nu p q} \eta_{\mu\sigma} +
\Gamma_{(3) \, \sigma p q} \eta_{\mu\nu} \right] ~. 
\end{eqnarray}
We use the convention that when a Lorentz index of $\Gamma_{(n)}^{\mu_1 \ldots
\mu_n}$ is contracted with an external momentum then that momentum appears in
place of the associated Lorentz index. The designation of $W_3$ in the 
superscript here indicates level. For $\partial W_3$ and $\partial \partial 
W_3$ the same basis is used. Given this choice we explicitly record the 
associated $14$~$\times$~$14$ matrix ${\cal M}^{W_3}$, \cite{10}. To save space
we set 
\begin{eqnarray}
{\cal M}^{W_3} &=& \frac{1}{2106d(d-2)} \left(
\begin{array}{cccc}
{\cal M}^{W_3}_{11} & {\cal M}^{W_3}_{12} & {\cal M}^{W_3}_{13} & 0 \\
{\cal M}^{W_3}_{21} & {\cal M}^{W_3}_{22} & {\cal M}^{W_3}_{23} & 0 \\
{\cal M}^{W_3}_{31} & {\cal M}^{W_3}_{32} & {\cal M}^{W_3}_{33} & 0 \\
0 & 0 & 0 & {\cal M}^{W_3}_{44} \\
\end{array}
\right) ~, \nonumber 
\label{matw3}
\end{eqnarray}
and emphasise that the $\Gamma_{(3)}$ sector corresponds to the lower outer 
corner submatrix or subspace. The clear partition is a reflection of the trace, 
(\ref{gamtr}). The explicit elements of each submatrix are  
\begin{eqnarray}
{\cal M}^{W_3}_{11} &=& \left(
\begin{array}{cccc}
312 (d+1) & 156 (d+1) & 78 (d+4) & 1248 (d+1) \\ 
156 (d+1) & 39 (5d+2) & 156 (d+1) & 624 (d+1) \\ 
78 (d+4) & 156 (d+1) & 312 (d+1) & 312 (d+4) \\ 
1248 (d+1) & 624 (d+1) & 312 (d+4) & 1664 (d+3) (d+1) \\ 
\end{array}
\right) ~, \nonumber \\ 
{\cal M}^{W_3}_{12} &=& \left(
\begin{array}{cccc}
\! 624 (d+1) & \! 312 (d+2) & 156 (d+4) & 624 (d+1) \\ 
\! 312 (2d+1) & \! 156 (3d+2) & 312 (d+1) & 312 (d+1) \\ 
\! 156 (3d+4) & \! 624 (d+1) & 624 (d+1) & 156 (d+4) \\ 
\! 832 (d+3) (d+1) & \! 416 (d+6) (d+1) & 208 (d+12) (d+1) & 832 (d+3) (d+1) \\ 
\end{array}
\right) ~, \nonumber \\ 
{\cal M}^{W_3}_{13} &=& \left(
\begin{array}{ccc}
624 (d+1) & 156 (3d+4) & 312 (d+4) \\ 
156 (3d+2) & 312 (2d+1) & 624 (d+1) \\ 
312 (d+2) & 624 (d+1) & 1248 (d+1) \\ 
416 (d+6) (d+1) & 208 (d+12) (d+1) & 104 (d^2+22d+48) \\ 
\end{array}
\right) ~, \nonumber \\ 
{\cal M}^{W_3}_{21} &=& \left(
\begin{array}{cccc}
624 (d+1) & 312 (2d+1) & 156 (3d+4) & 832 (d+3) (d+1) \\ 
312 (d+2) & 156 (3d+2) & 624 (d+1) & 416 (d+6) (d+1) \\ 
156 (d+4) & 312 (d+1) & 624 (d+1) & 208 (d+12) (d+1) \\ 
624 (d+1) & 312 (d+1) & 156 (d+4) & 832 (d+3) (d+1) \\ 
\end{array}
\right) ~, \nonumber \\ 
{\cal M}^{W_3}_{22} &=& (d+1) \left(
\begin{array}{cccc}
416 (2d+3) & 624 (d+2) & 104 \frac{(4d^2+25d+12)}{(d+1)} & 416 (d+3) \\ 
624 (d+2) & 208 \frac{(4d^2+7d+6)}{(d+1)} & 416 (2d+3) & 208 (d+6) \\ 
104 \frac{(4d^2+25d+12)}{(d+1)} & 416 (2d+3) & 416 (4d+3) & 104 (d+12) \\ 
416 (d+3) & 208 (d+6) & 104 (d+12) & 416 (4d+3) \\ 
\end{array}
\right) ~, \nonumber \\ 
{\cal M}^{W_3}_{23} &=& \left(
\begin{array}{ccc}
416 (d+3) (d+1) & 312 (d^2+7d+4) & 208 (d+12) (d+1) \\ 
312 (d+2)^2 & 416 (d+3) (d+1) & 416 (d+6) (d+1) \\ 
208 (d+6) (d+1) & 416 (d+3) (d+1) & 832 (d+3) (d+1) \\ 
416 (2d+3) (d+1) & 104 (4d^2+25d+12) & 208 (d+12) (d+1) \\ 
\end{array}
\right) ~, \nonumber \\ 
{\cal M}^{W_3}_{31} &=& \left(
\begin{array}{cccc}
624 (d+1) & 156 (3d+2) & 312 (d+2) & 416 (d+6) (d+1) \\ 
156 (3d+4) & 312 (2d+1) & 624 (d+1) & 208 (d+12) (d+1) \\ 
312 (d+4) & 624 (d+1) & 1248 (d+1) & 104 (d^2+22d+48) \\ 
\end{array}
\right) ~, \nonumber \\ 
{\cal M}^{W_3}_{32} &=& \left(
\begin{array}{cccc}
\! 416 (d+3) (d+1) \! & \! 312 (d+2)^2 \! & \! 208 (d+6) (d+1) & 
\! \! 416 (2d+3) (d+1) \! \\ 
\! 312 (d^2+7d+4) \! & \! 416 (d+3) (d+1) \! & \! 416 (d+3) (d+1) & 
\! \! 104 (4d^2+25d+12) \! \\ 
\! 208 (d+12) (d+1) \! & \! 416 (d+6) (d+1) \! & \! 832 (d+3) (d+1) & 
\! \! 208 (d+12) (d+1) \! \\
\end{array}
\right) \,, \nonumber \\ 
{\cal M}^{W_3}_{33} &=& \left(
\begin{array}{ccc}
208 (4d^2+7d+6) & 624 (d+2) (d+1) & 416 (d+6) (d+1) \\ 
624 (d+2) (d+1) & 416 (2d+3) (d+1) & 832 (d+3) (d+1) \\ 
416 (d+6) (d+1) & 832 (d+3) (d+1) & 1664 (d+3) (d+1) \\ 
\end{array}
\right) \,, \nonumber \\
{\cal M}^{W_3}_{44} &=& \left(
\begin{array}{ccc}
- 416 (d+1) & - 208 (d+1) & - 104 (d+4) \\
- 208 (d+1) & - 52 (5d+2) & - 208 (d+1) \\ 
- 104 (d+4) & - 208 (d+1) & - 416 (d+1) \\ 
\end{array}
\right) \,. 
\end{eqnarray}
These have been given in $d$-dimensions because we used dimensional 
regularization in our two loop computation.


\begin{thebibliography}{99}
\bibitem{1} C. Sturm, Y. Aoki, N.H. Christ, T. Izubuchi, C.T.C. Sachrajda \& 
A. Soni, Phys. Rev. {\bf D80} (2009), 014501.
%%CITATION = 0901.2599;%%
\bibitem{2} M. Gorbahn \& S. J\"{a}ger, Phys. Rev. {\bf D82} (2010), 114001.
%%CITATION = 1004.3997;%%
\bibitem{3} L.G. Almeida \& C. Sturm, Phys. Rev. {\bf D82} (2010), 054017.
%%CITATION = 1004.4613;%%
\bibitem{4} W. Celmaster \& R.J. Gonsalves, Phys. Rev. {\bf D20} (1979), 1420.
%%CITATION = PHRVA,D20,1420;%% 
\bibitem{5} G. Martinelli, C. Pittori, C.T. Sachrajda, M. Testa \& A. 
Vladikas, Nucl. Phys. {\bf B445} (1995), 81. 
%%CITATION = HEP-LAT 9411010;%% 
\bibitem{6} E. Franco \& V. Lubicz, Nucl. Phys. {\bf B531} (1998), 641
%%CITATION = HEP-LAT 9803491;%% 
\bibitem{7} K.G. Chetyrkin \& A. R\'{e}tey, Nucl. Phys. {\bf B583} (2000), 3. 
%%CITATION = HEP-PH 9910332;%% 
\bibitem{8} J.A. Gracey, Nucl. Phys. {\bf B662} (2003), 247.
%%CITATION = HEP-PH 0304113;%%
\bibitem{9} J.A. Gracey, JHEP {\bf 1103} (2011), 109.
%%CITATION = 1103.2055;%%
\bibitem{10} J.A. Gracey, Phys. Rev. {\bf D83} (2011), 054024.
%%CITATION = 1009.3895;%%
\bibitem{11} J.A. Gracey, Eur. Phys. J. {\bf C71} (2011), 1567.
%%CITATION = 1101.5266;%%
\bibitem{12} J.A. Gracey, JHEP {\bf 0904} (2009), 127.
%%CITATION = 0903.4623;%%
\bibitem{13} M. G\"{o}ckeler, R. Horsley, D. Pleiter, P.E.L. Rakow, 
A. Sch\"{a}fer and G. Schierholz, Nucl. Phys. Proc. Suppl. {\bf 119} (2003), 
32. 
%%CITATION = HEP-LAT 0209160;%% 
\bibitem{14} M. G\"{o}ckeler, R. Horsley, H. Oelrich, H. Perlt, D. Petters, 
P.E.L. Rakow, A. Sch\"{a}fer, G. Schierholz \& A. Schiller, Nucl. Phys. 
{\bf B544} (1999), 699.  
%%CITATION = HEP-LAT 9807044;%% 
\bibitem{15} S. Capitani, M. G\"{o}ckeler, R. Horsley, H. Perlt, P.E.L. Rakow,
G. Schierholz \& A. Schiller, Nucl. Phys. {\bf B593} (2001), 183.   
%%CITATION = HEP-LAT 0007004;%% 
\bibitem{16} C. Gattringer, M. G\"{o}ckeler, P. Huber \& C.B. Lang, Nucl. Phys.
{\bf B694} (2004), 170. 
%%CITATION = HEP-LAT 0404006;%% 
\bibitem{17} M. G\"{o}ckeler, R. Horsley, D. Pleiter, P.E.L. Rakow \& G.
Schierholz, Phys. Rev. {\bf D71} (2005), 114511. 
%%CITATION = HEP-PH 0410187;%% 
\bibitem{18} M. G\"{u}rtler, R. Horsley, P.E.L. Rakow, C.J. Roberts, G. 
Schierholz \& T. Streuer, PoS LAT2005 {\bf 125} (2006), 124. 
%%CITATION = HEP-LAT 0510045;%% 
\bibitem{19} M. G\"{o}ckeler, R. Horsley, Y. Nakamura, H. Perlt, D. Pleiter, 
P.E.L. Rakow, A. Sch\"{a}fer, G. Schierholz, A. Schiller, H. St\"{u}ben \& J.M.
Zanotti, Phys. Rev. {\bf D82} (2010), 114511.
%%CITATION = 1003.5756;%% 
\bibitem{20} J.B. Zhang, D.B. Leinweber, K.F. Liu \& A.G. Williams, Nucl. Phys.
Proc. Suppl. {\bf 128} (2004), 240. 
%%CITATION = HEP-LAT 0311030;%% 
\bibitem{21} D. Be\'{c}irevi\'{c}, V. Gimenez, V. Lubicz, G. Martinelli, M.
Papinutto \& J. Reyes, JHEP {\bf 0408} (2004), 022. 
%%CITATION = HEP-LAT 0401033;%% 
\bibitem{22} J.B. Zhang, N. Mathur, S.J. Dong, T. Draper, I. Horvath, F.X. Lee,
D.B. Leinweber, K.F. Liu \& A.G. Williams, Phys. Rev. {\bf D72} (2005), 
114509. 
%%CITATION = HEP-LAT 0507022;%% 
\bibitem{23} F. Di Renzo, A. Mantovi, V. Miccio, C. Torrero \& L. Scorzato, PoS
LAT2005 (2006), 237. 
%%CITATION = HEP-LAT 0509158;%% 
\bibitem{24} V. Gimenez, L. Giusti, F. Rapuano \& M. Talevi,  Nucl. Phys.
{\bf B531} (1998), 429.
%%CITATION = HEP-LAT 9806006;%% 
\bibitem{25} L. Giusti, S. Petrarca, B. Taglienti \& N. Tantalo, Phys. Lett.
{\bf B541} (2002), 350.
%%CITATION = HEP-LAT 0205009;%% 
\bibitem{26} A. Skouroupathis \& H. Panagopoulos, Phys. Rev. {\bf D79} (2009),
094508.
%%CITATION = 0811.4264;%%
\bibitem{27} M. Constantinou, P. Dimopoulos, R. Frezzotti, G. Herdoiza, K.
Jansen, V. Lubicz, H. Panagopoulos, G.C. Rossi, S. Simula, F. Stylianou \&
A. Vladikas, JHEP {\bf 1008} (2010), 068. 
%%CITATION = 1004.1115;%%
\bibitem{28} C. Alexandrou, M. Constantinou, T. Korzec, H. Panagopoulos \&
F. Stylianou, Phys. Rev. {\bf D83} (2011), 014503.
%%CITATION = 1006.1920;%%
\bibitem{29} R. Arthur \& P.A. Boyle, arXiv:1006.0422 [hep-lat].
%%CITATION = 1006.0422;%%
\bibitem{30} A.D. Kennedy, J. Math. Phys. {\bf 22} (1981), 1330.
%%CITATION = JMAPA,22,1330;%%
\bibitem{31} A. Bondi, G. Curci, G. Paffuti \& P. Rossi, Ann. Phys. {\bf 199}
(1990), 268.
%%CITATION = APNYA,199,268;%%
\bibitem{32} A.N. Vasil'ev, S.\'{E}. Derkachov \& N.A. Kivel, Theor. Math.
Phys. {\bf 103} (1995), 487.
%%CITATION = TMPHA,103,487;%%
\bibitem{33} A.N. Vasil'ev, M.I. Vyazovskii, S.\'{E}. Derkachov \& N.A. Kivel,
Theor. Math. Phys. {\bf 107} (1996), 441.
%%CITATION = TMPHA,107,441;%%
\bibitem{34} A.N. Vasil'ev, M.I. Vyazovskii, S.\'{E}. Derkachov \& N.A. Kivel,
Theor. Math. Phys. {\bf 107} (1996), 710.
%%CITATION = TMPHA,107,710;%%
\bibitem{35} J.A.M. Vermaseren, math-ph/0010025. 
%%CITATION = MATH-PH 0010025;%%
\bibitem{36} P. Nogueira, J. Comput. Phys. {\bf 105} (1993), 279. 
%%CITATION = JCTPA,105,279;%% 
\bibitem{37} S.G. Gorishny, S.A. Larin, L.R. Surguladze \& F.K. Tkachov,
Comput. Phys. Commun. {\bf 55} (1989), 381.
%%CITATION = CPHCB,55,381;%%
\bibitem{38} S. Laporta, Int. J. Mod. Phys. {\bf A15} (2000), 5087.
%%CITATION = HEP-PH 0207004;%%
\bibitem{39} C. Studerus, Comput. Phys. Commun. {\bf 181} (2010), 1293.
%%CITATION = 0912.2546;%%
\bibitem{40} C.W. Bauer, A. Frink \& R. Kreckel, cs/0004015.
%%CITATION = CS 0004015;%%
\bibitem{41} S.A. Larin \& J.A.M. Vermaseren, Phys. Lett. {\bf B303} (1993), 
334. 
%%CITATION = PHLTA,B303,334;%% 
\bibitem{42} J.C. Collins, {\it Renormalization} (Cambridge University Press,
1984).
\bibitem{43} A.I. Davydychev, J. Phys. {\bf A25} (1992), 5587.
%%CITATION = JPAGB,A25,5587;%%
\bibitem{44} N.I. Usyukina \& A.I. Davydychev, Phys. Atom. Nucl. {\bf 56}
(1993), 1553.
%%CITATION = HEP-PH 9307327;%%
\bibitem{45} N.I. Usyukina \& A.I. Davydychev, Phys. Lett. {\bf B332} (1994),
159.
%%CITATION = HEP-PH 9402223;%%
\bibitem{46} T.G. Birthwright, E.W.N. Glover \& P. Marquard, JHEP {\bf 0409}
(2004), 042.
%%CITATION = HEP-PH 0407343;%%
\bibitem{47} J.A. Gracey, JHEP {\bf 0610} (2006), 040. 
%%CITATION = HEP-PH 0609231;%%
\end{thebibliography}
\end{document}